\documentstyle[twoside,12pt]{article}
\begin{document}
\pagestyle{myheadings} \thispagestyle{empty} \normalsize
\noindent {\small G\"{o}teborg ITP 92-27, May 1992}\\
\begin{center}
\vspace*{20mm}
{\Large \bf Duality of string amplitudes in a curved background}\\
\vspace*{10mm} {\large M\aa ns Henningson}\\
\vspace*{5mm}
{\it Institute of Theoretical Physics, S-412 96  G\"{o}teborg, Sweden}\\
\vspace*{15mm}

\parbox{140mm}
{\small We initiate a program to study the relationship between the
target space, the spectrum and the scattering amplitudes in string
theory. We consider scattering amplitudes following from string theory
and quantum field theory on a curved target space, which is taken to
be the $SU(2)$ group manifold, with special attention given to the
duality between contributions from different channels. We give a
simple example of the equivalence between amplitudes coming from
string theory and quantum field theory, and compute the general form
of a four-scalar field theoretical amplitude. The corresponding string
theory calculation is performed for a special case, and we discuss how
more general string theory amplitudes could be evaluated.}
\end{center}

{\vspace*{10mm} \large \bf \noindent 1.  Introduction\\} The purpose
of this paper is to study the interplay between the geometry of the
target space of string theory, the string spectrum and the symmetry
properties of string scattering amplitudes. We will consider string
theory and quantum field theory in parallel. In the following we give
some reasons why we hope that such a program might be fruitful.

String theory is a remarkably unique theory. In addition to the
bosonic string, the Ramond-Neveu-Schwarz string and the Green-Schwarz
string, one may also mention the heterotic string, which is really a
synthesis of the Ramond-Neveu-Schwarz string and the bosonic string,
but this more or less exhausts the list. Furthermore, the
Ramond-Neveu-Schwarz string and the Green-Schwarz string are probably
equivalent, although formulated in rather different ways. It is true
that we have a plethora of (perturbative) vacua, but there are reasons
to believe that non-perturbative tunneling effects will eventually
select a unique one, or at least narrow down the choice considerably.
Anyway, these different vacua should be regarded as different
solutions to the same equations, rather than as distinct theories.

We may get a hint of why string theory is so unique already from a
popular account of the theory. The original string action for the
freely propagating string was simply proportional to the area of the
world-sheet swept out by the string \cite{Nambu}\cite{Goto}, and thus
has a strong flavour of geometry. This is even more apparent when we
consider interactions. In point particle theories, we have
distinguished interaction vertices where world-lines meet. Nothing
such is possible in string theory, since we cannot give a Lorentz
invariant definition of where the interaction takes place. The
interactions of a string theory are therefore more or less determined
by the propagation of a free string, and we have few coupling
constants to adjust. Finally, the quantum mechanical consistency of a
string theory seems to be a very delicate issue with many potential
anomalies. The proper cancelation of such anomalies puts severe
restrictions on the choice of symmetry groups and representations.

Symmetry arguments (in flat space) usually relate particles of the
same mass and spin (for example the gauge group of the standard model
$SU(3) \times SU(2) \times U(1)$ or flavour $SU(3)$), or exceptionally
particles of the same mass but different spins (supersymmetry). String
theory seems to go beyond these limitations in that it assembles
particles of different masses in multiplets, as in the familiar
spectra of the various string theories. If we regard such a spectrum
as an ``irreducible representation'' of a hitherto unknown ``string
symmetry group'', it is plausible that all scattering amplitudes are
related to each other. This suggests that we regard string theory in
the following way: The in-data supplied is the target space in which
the string lives (for example Minkowski space, superspace, a group
manifold,...), a parameter related to the size of a typical string
(the string tension, the level,...) and a single string coupling
constant. The usual axioms for an $S$-matrix theory, together with
some unknown ``string principle'' should then determine not only the
spectrum of physical states of the theory, but also the amplitudes for
scattering of those states. Furthermore, it seems that the consistency
of string theory puts severe restrictions on the allowed target
spaces.

String theory dates its roots back to the days of $S$-matrix theory.
The first result in what was to become string theory was the Veneziano
amplitude \cite{Veneziano}, which was thought to be the Born term of a
four-point amplitude for strongly interacting particles. An
alternative amplitude was proposed in \cite{Virasoro}. These
amplitudes attracted interest because of a remarkable symmetry
property, called duality \cite{Dolen-Horn-Schmid}, which relates
contributions in different channels. It was only somewhat later that
it was realized that these amplitudes may be given an interpretation
in terms of a theory of relativistic strings. The trend in the last
few years has been to focus more on string theory as a two-dimensional
theory defined on the world-sheet swept out by the propagating string,
and less on its target space properties. This approach has been
remarkably fruitful, but several important questions remain
unanswered. It is a common belief that a deep understanding of string
theory would require a completely new formulation. To find such a
formulation, we should scrutinize string theory from as many different
perspectives as possible.

The great interest in string theory stems largely from the hope that
it will prove a viable way to quantum gravity
\cite{Yoneya}\cite{Scherk-Schwarz}. Indeed, the string spectrum
contains states which may be identified as gravitons, and string
theory seems to provide a consistent scheme for calculating
perturbative graviton-graviton scattering amplitudes. However, most
research in string theory concerns strings propagating in a flat
Minkowski space, but if we are to describe gravity, we must also
consider more general backgrounds. A truly consistent theory must
allow for the string to influence its own background, but this is as
yet beyond our understanding. Anyway, a study of strings in any
background different from flat Minkowski space should be worthwhile.

Minkowski space is in many respects the simplest possible background
for string theory, but for a study of these questions it is not so
well suited.  The group theory of the underlying isometry group, the
Poincar\'{e} group, is rather involved. We hope that something could
be learnt by studying simpler homogeneous spaces. Most of this paper
will be devoted to bosonic strings on target manifolds which could be
equipped with a group structure, in particular $SU(2)$. Our reason for
this is largely technical. We feel that there is good hope, though,
that the features we are interested in will survive even in such an
extremly simplified and unrealistic toy model.

We have seen that string theory has departed rather much from its
origin. Our intention is to focus on the properties of the final
result, the string amplitudes, rather than on intermediate steps in
the calculations. This is in the spirit of the old dual models ideas.
A comparatively large part of this paper is devoted to a general
discussion of scattering amplitudes from string theory and quantum
field theory, but we also give a few concrete examples to illustrate
the ideas. We hope to be able to present more realistic examples in
forthcoming publications.

This paper is organized as follows: In section~2 we give a brief
review of some aspects of string theory in a curved background.
Section~3 is devoted to explain how a scattering process in symmetric
spaces is described. In section~4 we discuss the calculation of
scattering amplitudes in a quantum field theory. This discussion is
specialized to quantum field theory on a group manifold in section~7
after a review of some group theory in section~5 and a discussion of
the string spectrum in section~6. In sections 8, 9 and 10 we give some
examples of how string theory amplitudes could be computed. In
section~11 we briefly consider the flat space limit, and finally, in
section~12, we discuss the relevance of the present work, and indicate
how we intend to continue the programme.

{\vspace*{10mm} \large \bf
\noindent 2.  String theory in a curved background\\}
String theory is most often treated in a first quantized formalism,
i.~e. as a two-dimensional quantum field theory defined on the
world-sheet of the string. For the string interpretation to be
consistent, it is necessary that this quantum field theory is
invariant, not only under general reparametrizations of the world
sheet, but also under scale changes. It should thus be a conformal
field theory.

We will restrict our attention to a purely bosonic string moving on a
$D$-dimensional target manifold $\cal M$. If we introduce (local)
coordinates $x^\mu$, $\mu = 1,\ldots,D$ on $\cal M$, we may write down
the most general renormalizable string action:
\begin{eqnarray}
S & = & \frac{1}{\alpha^\prime} \int d^2 \sigma \; \sqrt{g}
g^{\alpha\beta} \partial_\alpha x^\mu \partial_\beta x^\nu G_{\mu \nu}
(x) \\ & & + \frac{1}{\alpha^\prime} \int d^2 \sigma \;
\epsilon^{\alpha \beta} \partial_\alpha x^\mu \partial_\beta x^\nu
B_{\mu \nu} (x) + \int d^2 \sigma \; \sqrt{g} R^{(2)} \phi (x).
\nonumber
\end{eqnarray}
Here $g^{\alpha \beta}$ and $R^{(2)}$ denote the metric and the
curvature scalar of the world-sheet respectively, and $\alpha^\prime$
is a constant of (target space) dimension $[{\rm length}]^2$. The
functions $G_{\mu \nu} (x)$, $B_{\mu \nu} (x)$ and $\phi (x)$ are
interpreted as a metric, an anti-symmetric tensor field and a scalar
field (the dilaton) on the target space $\cal M$ respectively.

The requirement that the theory be conformally invariant at quantum
level amounts to the vanishing of the beta-functionals of the
couplings $G_{\mu \nu} (x)$ and $B_{\mu \nu} (x)$. To the first
non-vanishing order in $\alpha^\prime$ this means
\cite{Callan-Friedan-Martinec-Perry} that
\begin{eqnarray}
0 & = & R_{\mu \nu} + \frac{1}{4} H_\mu^{\;\;\lambda \rho} H_{\nu
\lambda \rho} - 2 D_\mu D_\nu \phi \label{eq2.2}\\ 0 & = & D_\lambda
H^\lambda_{\;\;\mu \nu} - 2 ( D_\lambda \phi ) H^\lambda_{\;\;\mu
\nu}, \nonumber
\end{eqnarray}
where $H_{\mu \nu \rho} = 3 D_{[\mu} B_{\nu \rho]}$ and $R_{\mu \nu}$
is the Ricci tensor corresponding to the metric $G_{\mu \nu} (x)$.

A truly consistent string theory also requires the conformal anomaly
$c$, including contributions from the Fadeev-Popov ghosts that arise
upon gauge fixing, to vanish. This means that the beta-functional of
$\phi (x)$ should vanish, which to lowest non-trivial order in
$\alpha^\prime$ yields
\begin{equation}
0 = \frac{D-26}{3 \alpha^\prime} + 4 D_\mu \phi D^\mu \phi - 4 D_\mu
D^\mu \phi + R + \frac{1}{12} H_{\mu \nu \rho} H^{\mu \nu \rho}.
\end{equation}
However, we will only consider string theory at genus zero (tree
level), where the conformal anomaly is of no consequence. The toy
model that we will describe later in fact has a negative conformal
anomaly. The reader who feels uneasy about this may always add other
conformal field theories, such as for example free bosons or minimal
models, to make up for the missing anomaly.

It is obviously not easy to find solutions to the non-linear equations
(\ref{eq2.2}), especially if we include terms of higher orders in
$\alpha^\prime$. An interesting possibility would be a string moving
in a maximally symmetric space so that $R_{\mu \nu \rho \sigma} \sim
g_{\mu \rho} g_{\nu \sigma} - g_{\mu \sigma} g_{\nu \rho}$. As we will
see in the next section, the description of scattering is faciliated
in a maximally symmetric space. However, it should be realized that
the fields $B_{\mu \nu}$ and $\phi$ will in general break this
symmetry. Namely, if we use the condition that $H_{\mu \nu \rho}$ is a
maximally symmetric tensor (see for example \cite{Helgason}),
\begin{equation}
\delta^\tau_\mu H^\sigma_{\;\; \nu \rho} + \delta^\tau_\nu
H_{\mu \;\; \rho}^{\;\; \sigma} + \delta^\tau_\rho H_{\mu \nu}^{\;\;\;\;
\sigma} = \delta^\sigma_\mu H^\tau_{\;\; \nu \rho} + \delta^\sigma_\nu
H_{\mu \;\; \rho}^{\;\; \tau} + \delta^\sigma_\rho H_{\mu \nu}^{\;\;\;\; \tau},
\end{equation}
and contract with $\delta^\mu_\tau$, we see that we must require that
$D=3$ (so that $H_{\mu \nu \rho} \sim \epsilon_{\mu \nu \rho}$).

An important class of solutions to (\ref{eq2.2}) have $\phi \equiv 0$
and $B_{\mu \nu}$ chosen so that $H_{\mu \nu \rho}$ acts as a
parallelizing torsion \cite{Curtright-Zachos}. Such torsions only
exist for manifolds which admit a group structure and for $S_7$ with
the round metric \cite{Cartan-Schouten}\cite{Einstein}.  The latter
possibility is excluded, however, since the parallelizing torsion on
$S_7$ is non-closed and therefore non-exact (even locally).

Although the corrections to (\ref{eq2.2}) are not known to arbitrarily
high orders in $\alpha^\prime$, it has been shown that a group
manifold with a parallelizing torsion is an exact solution
\cite{Mukhi}. This is in fact not too astonishing. These models are
the familiar Wess-Zumino-Witten models
\cite{Witten}\cite{Knizhnik-Zamolodchikov}\cite{Gepner-Witten}. They
are exactly solvable conformal field theories, and thus fixed points
of the renormalization group.

A Wess-Zumino-Witten model is completely specified by its symmetry
group $G$ and an integer $k$ (the level), which is inversely
proportional to $\alpha^\prime$. We will be mostly concerned with
$G=SU(2)$. This group is three-dimensional, and the metric is
maximally symmetric. Models based on the closely related group
$SU(1,1)$ have attracted much interest recently
\cite{Hwang}\cite{Witten2}. They offer the attractive feature of
containing a time-direction, and therefore seem closer to physical
reality. Many questions concerning the $SU(1,1)$ Wess-Zumino-Witten
model still remain unanswered, though. We will not consider $SU(1,1)$
in this paper, but we hope that a thorough understanding of $SU(2)$
string amplitudes will prove helpful when investigating other string
theories in general and $SU(1,1)$ strings in particular. Namely, if
the final result is expressible in purely group theoretical terms, it
should be possible to translate it from one group to another.

Strictly speaking, an $S$-matrix theory presupposes that the external
states are free, i.~e. the fields should obey a free field equation at
infinity. On a compact manifold, such as the $SU(2)$ group manifold,
we cannot construct such asymptotic states, and one might therefore
argue that the whole idea of scattering is ill defined. What we are
really calculating in this paper, however, is amputed amplitudes or
Green functions. It seems that our prescription for calculating such
Green functions is consistent, and this is sufficient for our
purposes. We stress once more that we are considering a toy model,
which in many respects is far from physical reality, but that our
primary object is to investigate certain properties of its scattering
amplitudes.

{\vspace*{10mm} \large \bf \noindent 3.  Kinematics in a curved
space\\}
In this section we will investigate the constraints on
particle scattering amplitudes that follow from the geometry of the
target space.

The particle concept is in general not well defined in an arbitrarily
curved space. (See for example \cite{Birrell-Davies}.) We will
consider a theory invariant under some symmetry group $\cal G$, which
may include internal symmetries in addition to space-time isometries,
and we will use Wigner's definition of an elementary particle: An
elementary system corresponds to a unitary representation of ${\cal
G}$. This procedure is correct at least if the space is maximally
symmetric.

To a particle transforming in the $D_R$ representation under $\cal G$
there is associated a representation space $H_R$. Consider now a
scattering situation with $n$ external particles transforming in the
$D_{R_1},\ldots,D_{R_n}$ representations. This configuration
transforms in the tensor product representation $D_{tot} = D_{R_1}
\otimes \ldots \otimes D_{R_n}$, defined in the space $H_{tot} =
H_{R_1} \otimes \ldots \otimes H_{R_n}$, under $\cal G$.

A scattering amplitude $\Gamma$ could now be regarded as a linear
functional on this space $H_{tot}$. We should require $\Gamma$ to be
invariant under $\cal G$. If we decompose $D_{tot}$ as a direct sum of
irreducible representations, it is not difficult to see that this
amounts to the vanishing of $\Gamma$ on all subspaces of $H_{tot}$
which transform non-trivially under $\cal G$. The values of $\Gamma$
on the subspace $H_0$ of $\cal G$ invariant states are arbitrary. The
scattering amplitude $\Gamma$ is thus completely specified by its
values on a set of basis vectors of $H_0$.

This way of viewing scattering may seem unfamiliar, and we will
therefore briefly consider the well-known example when $\cal G$ is the
four-dimensional Poincar\'{e} group. Its unitary irreducible
representations are characterized by a mass and a spin \cite{Wigner}.
We will take the external particles to be spinless and of mass $m$, so
that the vectors $| p^\mu>$ where $p^2=m^2$ constitute a basis for the
one-particle space. In the case of four external particles the space
$H_{tot}$ is thus spanned by the vectors
\begin{equation}
|p_1,\ldots,p_4> = |p_1> \otimes \ldots \otimes |p_4>,
\end{equation}
where $p_1^2 = \ldots = p_4^2 = m^2$, and a general state in $H_{tot}$
may be written as
\begin{equation}
|f> = \prod_{i=1}^4 \int d^4 p_i \; \delta (p_i^2 - m^2)
f(p_1,\ldots,p_4) |p_1,\ldots,p_4>
\end{equation}
for an arbitrary function $f(p_1,\ldots,p_4)$. The requirement that
$|f>$ is $\cal G$ invariant is equivalent to demanding that
\begin{equation}
f(p_1,\ldots,p_4)=\tilde{f}((p_1+p_2)^2,(p_1+p_3)^2) \delta^{(4)}
(p_1+\ldots + p_4)
\end{equation}
for some function $\tilde{f}(s,t)$. As a basis for $H_0$ we may take
the vectors
\begin{eqnarray}
|s,t> & = & \prod_{i=1}^4 \int d^4 p_i \; \delta (p_i^2-m^2)
\delta((p_1+p_2)^2- s) \delta((p_1+p_3)^2-t) \\
& & \delta^{(4)}(p_1+\ldots +p_4) |p_1,\ldots,p_4> \nonumber,
\end{eqnarray}
and the scattering amplitude $\Gamma$ could be described by its values
$\Gamma (s,t)$ on these vectors. We have thus recovered the usual
description of a four-point amplitude in terms of the Mandelstam
variables $s$ and $t$.

Let us now return to a general symmetry group $\cal G$. The
Clebsch-Gordan series for its representations provides us with a
suitable basis for the subspace $H_0$ of $\cal G$ invariant states, as
we will now explain. The tensor product of two irreducible
representations is in general reducible, and may be decomposed as a
direct sum of irreducible representations. We will assume that to
every irreducible representation $D_R$ there is a conjugate
irreducible representation $D_{\bar{R}}$ so that the tensor product of
$D_R$ and $D_{\bar{R}}$ contains exactly one copy of the trivial
one-dimensional representation.  Furthermore, we will assume that the
tensor product of $D_R$ with any other irreducible representation does
not contain the trivial representation. As a basis for $H_0$ for the
scattering of four external particles transforming in the
$D_{R_1},\ldots,D_{R_4}$ representations we may now take the
orthonormal vectors $|R>$, which transform trivially under $\cal G$
acting on all four external particles, and belong to the
representation space of $D_R$ in the Clebsch-Gordan decomposition of
$D_{R_1} \otimes D_{R_2}$. Our scattering amplitude $\Gamma$ may thus
be described by a function $\Gamma (R)$, where $R$ runs over a set of
representations of $\cal G$. By chosing a different pairing of the
external particles we may obtain a different basis. The matrix which
relates the bases is the $\cal G$ counterpart of the Wigner $6j$
symbol for $SU(2)$.

If two of the external particles are identical, the operation of
permuting them is a well-defined map from $H_0$ to itself, and thus
also from the dual space $H_0^*$ of scattering amplitudes to itself.
It thus makes sense to say that an amplitude is symmetric under
permutation of those external particles.  An analogous reasoning
applies in the case that all four external particles are identical.
Such ``crossing symmetry'' is in fact one of the requirements that is
usually imposed on an $S$-matrix theory
\cite{Eden-Landshoff-Olive-Polkinghorne}.

{\vspace*{10mm} \large \bf \noindent 4.  Field theoretical scattering
amplitudes\\}
Sofar we have only used ``kinematical'' symmetry
arguments to determine the possible form of the scattering amplitude
$\Gamma$. We will now impose the ``dynamical'' constraint that
$\Gamma$ follows from a local quantum field theory.

Let us assume that the symmetry group $\cal G$ is the isometry group
of the target space manifold $\cal M$. This means that an element of
$\cal G$ acts as a metric preserving map from $\cal M$ to itself.
These maps induce linear transformation laws for tensor fields on
$\cal M$. Each field carries a reducible representation of $\cal G$,
which may be decomposed as a direct sum of irreducible components.

In a Lagrangian formalism, the theory is defined by a set of such
fields $\phi^i(x)$ and an action functional $S[\phi]$, which should be
$\cal G$ invariant. The action $S[\phi]$ is often decomposed as a sum
of a kinetic term $S_{kin}$, which is bilinear in the fields
$\phi^i(x)$, and an interaction term $S_{int}$, which is trilinear or
higher. If we denote the part of $\phi^i(x)$ that transforms in the
$D_R$ representation under $\cal G$ as $\phi^i_R$, we may write
\begin{equation}
S_{kin} = \sum_i \sum_R \phi^i_R \phi^i_{\bar{R}}
C_{R\bar{R}}K_i(R)\label{eq4.1}
\end{equation}
and
\begin{equation}
S_{int} = \sum_{ijk} \sum_{R_i R_j R_k} \phi^i_{R_i} \phi^j_{R_j}
\phi^k_{R_k} C_{R_i R_j R_k} V_{ijk} (R_i,R_j,R_k) + {\cal
O}(\phi^4),\label{eq4.2}
\end{equation}
where $C_{R\bar{R}}$ ($C_{R_i R_j R_k}$) denotes the Clebsch-Gordan
coefficient for coupling two (three) representations of $\cal{G}$ to
yield the trivial representation. If a certain representation does not
occur in $\phi^i(x)$, or if the three representations may not be
coupled together, we will assume that the corresponding $V_{ijk}
(R_i,R_j,R_k)$ vanishes.

We should now calculate the tree-level contribution to the scattering
amplitude for four external particles transforming in the
$R_1,\ldots,R_4$ representations. In a path integral quantization it
is easy to see what happens.  The amplitude is a sum of contributions
in different channels:

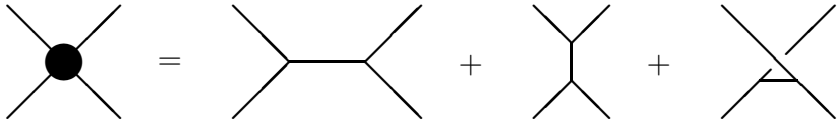
\setlength{\unitlength}{0.5mm} \thicklines
\begin{picture}(250,40)(-30,0) \put(0,0){\line(1,1){30}}
\put(0,30){\line(1,-1){30}} \put(15,15){\circle*{10}} \put(40,13){=}
\put(60,0){\line(1,1){15}} \put(60,30){\line(1,-1){15}}
\put(75,15){\line(1,0){20}} \put(95,15){\line(1,1){15}}
\put(95,15){\line(1,-1){15}} \put(120,12){+} \put(140,0){\line(1,1){10}}
\put(140,30){\line(1,-1){10}} \put(150,10){\line(0,1){10}}
\put(150,10){\line(1,-1){10}} \put(150,20){\line(1,1){10}} \put(170,12){+}
\put(190,0){\line(1,1){13}} \put(207,17){\line(1,1){13}}
\put(190,30){\line(1,-1){30}} \put(200,10){\line(1,0){10}}
\end{picture}

To calculate a diagram we need the vertex factor $V_{ijk}
(R_i,R_j,R_k)$ and the propagator $K^{-1}_i(R)$. The rest is pure
group theory. It is not difficult to see that the $s$-channel
contribution to the function $\Gamma (R)$ introduced in the last
section is
\begin{equation}
\Gamma^{(s)}_{i_1\ldots i_4}(R) = \sum_i V_{i_1 i_2 i} (R_1,R_2,R) K^{-1}_i
(R) V_{i i_3 i_4} (R,R_3,R_4).
\end{equation}
The contributions from the other channels are most easily calculated
in the corresponding bases of $H_0$ and then transformed to the
$s$-channel basis by means of the ``$6j$'' symbols.

If the external particles are identical, the total amplitude is
obviously symmetric under permutation of the external legs, as we
described in the last section. However, we may conceive of a theory in
which already the $s$-channel contribution is symmetric under exchange
of two external particles (planar duality), or the sum of the $s$- and
$t$-channel contributions is symmetric under permutation of all four
external particles (non-planar duality). This is clearly a non-trivial
constraint on the field content and the couplings of the field theory.

There are good reasons to believe that duality is an essential
property of string theory, and therefore a plausible candidate for a
``string principle''. In fact, there is no clear distinction between
diagrams in different channels in string theory. Using local conformal
rescalings of the world-sheet metric, we may relate seemingly
different diagrams.

The functions $K_i (R)$ and $V_{i_1 i_2 i_3} (R_1,R_2,R_3)$ may not be
chosen at will, but are constrained by locality of the quantum field
theory. To see how this works we will first consider the simplest
case, namely a scalar field $\phi (x)$. The most general kinetic term
is
\begin{equation}
S_{kin} = \int d^D x \; \phi (x) ( \Box - \mu^2) \phi (x), \label{eq4.4}
\end{equation}
where $\Box$ is the d'Alembertian on $\cal M$ and $\mu$ is a constant.
We have assumed at most two derivatives in analogy with flat space
field theory, where terms with more derivatives lead to non-unitary
theories. As we have already mentioned, $\phi (x)$ may be decomposed
into its irreducible components $\phi_R$. The d'Alembertian is a
Casimir operator of $\cal G$, so the components $\phi_R$ are
eigenfunctions of $\Box$ with some eigenvalue $Q(R)$. The
decomposition of $\phi$ into $\phi_R$ thus amounts to doing harmonic
analysis on $\cal M$. The kinetic term (\ref{eq4.4}) may now be
rewritten in the form (\ref{eq4.1}) with $K_\phi(R)=Q(R)-\mu^2$.

We should now scatter external $\sigma(x)$ fields by the exchange of
$\phi(x)$ quanta. The only trilinear $\cal G$ invariant interaction
term without any derivatives is
\begin{equation}
S_{int} = \int d^D x \; \lambda \phi(x) \sigma(x) \sigma(x).
\end{equation}
With the same normalization of the irreducible field components as in
the kinetic term this corresponds to an interaction of the form
(\ref{eq4.2}) with $V_{\sigma \sigma \phi} (R_1,R_2,R) = 1$. According
to our previous results the $s$-channel contribution to the four point
amplitude is thus $\Gamma (R) = (Q(R)-\mu^2)^{-1}$.

We could now in principle go on and calculate the contributions from
exchange of vector fields $A_\mu (x)$ and higher rank tensor fields
$A_{\mu_1 \ldots \mu_r} (x)$. However, the decomposition of the fields
into their irreducible components and the determination of how these
representations are coupled together in the action in general require
a great deal of knowledge about the differential geometry of the
manifold $\cal M$.

{\vspace*{10mm} \large \bf \noindent 5.  Differential geometry on
group manifolds\\}
To write down our Lagrangian in a covariant way, so
that the general covariance is manifest, we need a machinery for doing
tensor algebra on the target space. The actual evaluation of the
theory is most often performed in Fourier space, where the derivatives
in the Lagrangian are simply algebraic operations. The interpretation
of harmonic eigenfunctions as asymptotic states is also easier in
Fourier space. Both of these technical questions, tensor algebra and
harmonic analysis, are much simpler if the target space may be
equipped with a group structure. As is well known, there is a close
relationship between harmonic analysis on a group manifold and the
representation theory of the group. For tensor analysis, we have a
canonical choice for a vielbein, namely the left (or right) invariant
vector fields associated with the Lie algebra of the group.

Let us therefore assume $\cal M$ to be the group manifold of some Lie
group $G$. The isometry group is then ${\cal G} = G \otimes G$, and
acts as
\begin{equation}
{\cal G} \ni (u,v): \;\;\;\; {\cal M} \ni x \mapsto x^\prime =
uxv^{-1} \in {\cal M}. \label{eq5.1}
\end{equation}

Anticipating that $G = SU(2)$, we will denote the unitary irreducible
representations of $G$ as $D_j$, and label the states within such a
representation as $|j,m>$. A $\cal G$ representation
$D_R=(D_{j_L},D_{j_R})$ is then specified by giving the
representations $D_{j_L}$ and $D_{j_R}$ of the left and right $G$
factor.

We will often have reason to consider different ways of coupling
representations together. Such recouplings are described by the Wigner
$3nj$ symbols for the case $G = SU(2)$. In particular we will often
use the formula
\begin{equation}
\begin{array}{c} {\small \begin{picture}(30,30)  \put(5,5){\line(1,2){5}}
\put(5,25){\line(1,-2){5}} \put(20,15){\line(1,-2){5}}
\put(20,15){\line(1,2){5}} \put(10,15){\line(1,0){10}} \put(0,27){$j_1$}
\put(26,27){$j_2$} \put(0,1){$j_3$} \put(26,1){$j_4$}  \put(13,9){$j$}
\end{picture} } \end{array} = \sum_{j^\prime}
(\!-\!1)^{j_2\!+\!j_3\!+\!j\!+\!j^\prime} [(2j+1)(2j^\prime+1)]^{\frac{1}{2}}
\left\{ \begin{array}{ccc} j_1&j_2&j\\j_4&j_3&j^\prime \end{array} \right\}
\begin{array}{c} {\small \begin{picture}(30,30)  \put(5,5){\line(2,1){10}}
\put(5,25){\line(2,-1){10}} \put(15,10){\line(2,-1){10}}
\put(15,20){\line(2,1){10}} \put(15,10){\line(0,1){10}}  \put(0,27){$j_1$}
\put(26,27){$j_2$} \put(0,1){$j_3$} \put(26,1){$j_4$} \put(9,12){$j^\prime$}
\end{picture} } \end{array}, \label{15}
\end{equation}
which describes the relation between different ways of coupling
$D_{j_1}, \ldots, D_{j_4}$ to yield the trivial representation. (See
for example \cite{Varshalovich}.)

It is well-known that the elements $V^j_{m\bar{m}} (x) =
<j,m|V^j(x)|j,\bar{m}>$ of the representation matrices form an
orthogonal basis of ${\cal L}_2 (G)$, or explicitly
\begin{equation}
\int d \mu (x) \; V^j_{m\bar{m}}(x)V^{j^\prime}_{m^\prime \bar{m}^\prime} (x)
= \delta_{jj^\prime}\delta_{mm^\prime}\delta_{\bar{m}\bar{m}^\prime}\frac{1}
{{\rm dim}\; D_j},
\end{equation}
where $d \mu (x)$ is the $\cal G$ invariant Haar measure on $G$. (See
for example \cite{Barut-Raczka}.) We may thus expand a scalar field
$\phi(x)$ on $G$ as
\begin{equation}
\phi(x)=\sum_j\sum_{m\bar{m}} \phi^j_{\bar{m}m} V^j_{m\bar{m}}(x),
\end{equation}
where the Fourier components $\phi^j_{m\bar{m}}$ are given by
\begin{equation}
\phi^j_{m\bar{m}} = {\rm dim}\;D_j \int d\mu(x) \; \phi(x)
\bar{V}^j_{\bar{m}m}(x).
\end{equation}
It is not difficult to verify that the scalar field transformation law
for $\phi(x)$ under the isometry group implies that
$\phi^j_{\bar{m}m}$ transforms in the $(D_j,D_{\bar{\jmath}})$
representation under $\cal G$ acting as in (\ref{eq5.1}). Here
$D_{\bar{\jmath}}$ denotes the conjugate representation of $D_j$. The
full representation content of a scalar field $\phi(x)$ is thus
\begin{equation}
D_0 = \bigoplus_j (D_j,D_{\bar{\jmath}}),\label{eq5.5}
\end{equation}
where the sum runs over all unitary representations of $G$.

Before we treat vector fields and higher rank tensor fields we need to
introduce some more notation. A group element $x \in G$ may be
expanded as $x=\exp(i\theta_a T^a)$, where the $T^a$ span the Lie
algebra $g$ of $G$ and fulfil
\begin{equation}
[T^a,T^b] = if^{ab}_{\;\;\;\;c} T^c.\label{eq5.6}
\end{equation}
Two especially important solutions to these commutation relations are
the trivial one-dimensional representation $D_{triv}$, in which
$T^a_{triv}=0$, and the ${\rm dim}\;G$ dimensional adjoint
representation $D_{adj}$, in which
\begin{equation}
(T^a_{adj})^b_{\;\;c} = -i f^{ab}_{\;\;\;\;c}.\label{eq5.6b}
\end{equation}

We introduce a metric $\eta^{ab}$ in $g$ through $\eta^{ab}={\rm
Tr}(T^a T^b)$, where ${\rm Tr}$ denotes the (suitably normalized)
trace in an arbitrary representation. Lie algebra indices are raised
and lowered with $\eta^{ab}$ and its inverse $\eta_{ab}$.

It is straightforward to verify that the matrices $T^a$ in the $D_j$
representation are proportional to the Clebsch-Gordan coefficients for
coupling of $D_j$ and $D_{adj}$ to yield $D_j$, or more precisely
\begin{equation} (T^a_j)^m_{m^\prime} = \sqrt{Q(j)}
C^{adj\;j\;j}_{\;\;a\;\;m\;m^\prime}. \label{eq5.7}
\end{equation}
Here $Q(j)$ denotes the eigenvalue of the quadratic Casimir operator
$Q=T^aT_a$ in the $D_j$ representation.

To treat tensorfields we introduce a set of (local) coordinates
$x^\mu$, \linebreak $\mu =1,\ldots,{\rm dim}\;G$ on $\cal M$ and
define the vielbein $e^a_\mu(x)$ as
\begin{equation}
e^a_\mu(x)= {\rm Tr}(T^a\partial_\mu x x^{-1}).
\end{equation}
Here $x$ denotes an element of $G$, or equivalently a point in $\cal
M$. The $\cal G$ invariant metric on $\cal M$ is given by
\begin{equation}
g_{\mu \nu}(x)=\eta_{ab}e^a_\mu(x)e^b_\nu(x),
\end{equation}
and has inverse $g^{\mu\nu}(x)=e_a^\mu(x) e_b^\nu(x) \eta^{ab}$, where
$e_a^\mu(x)$ is the inverse of $e^a_\mu(x)$. The Haar measure could
now be written as $d\mu(x) = d^D x \sqrt{\det g_{\mu\nu}}$. Using
\begin{equation}
\partial_\mu e^a_\nu-\partial_\nu e^a_\mu = if^{abc} e_{b\mu} e_{c\nu}
\end{equation}
we may calculate the affine connection
\begin{eqnarray}
\Gamma^\lambda_{\mu \nu} & = & \frac{1}{2} g^{\lambda \rho}
(\partial_\mu g_{\nu \rho}+\partial_\nu g_{\mu \rho} - \partial_\rho g_{\mu
\nu}) = \frac{1}{2} e_a^\lambda (\partial_\mu e_\nu^a + \partial_\nu e^a_\mu)
\end{eqnarray}
and the covariant derivative
\begin{equation}
D_\mu e^a_\nu = \partial_\mu e^a_\nu - \Gamma^\lambda_{\mu\nu}
e^a_\lambda = \frac{i}{2} f^{abc} e_{b \mu} e_{c \nu}.
\end{equation}
Finally, we note that
\begin{equation}
e^{a\mu}(x) D_\mu V^j_{m\bar{m}}(x) = \sum_n (T^a_j)_{mn}
V^j_{n\bar{m}}(x), \label{eq5.13}
\end{equation}
and that
\begin{equation}
D_\mu D^\mu V^j_{m\bar{m}}(x) = Q(j) V^j_{m\bar{m}}(x).
\end{equation}

The vielbein $e^a_\mu(x)$ should transform as a covariant vector under
the isometry (\ref{eq5.1}), i.~e. as
\begin{equation}
e^a_\mu(x) \mapsto e^{\prime a}_\mu(x^\prime) = \frac{\partial
x^\nu}{\partial x^{\prime\mu}} e^a_\nu(x) = (V^{adj}(u))^a_{\;\;b}
e^b_\mu(x),
\end{equation}
where $(V^{adj}(u))^a_{\;\;b} = {\rm Tr}(uT^au^{-1}T_b)$ is the
representation matrix of $u$ in the adjoint representation. The
vielbein $e^a_\mu(x)$ thus transforms as $(D_{adj},D_{triv})$ under
${\cal G}=G\otimes G$ acting as in (\ref{eq5.1}). Equivalently, we
could have worked with $\tilde{e}^a_\mu(x)= {\rm Tr}(T^a x^{-1}
\partial_\mu x)$, which transforms as $(D_{triv},D_{adj})$.

A general tensorfield $A_{\mu_1\ldots\mu_r}(x)$ may now be expanded as
\begin{equation}
A_{\mu_1\ldots\mu_r}(x) = e^{a_1}_{\mu_1}(x) \ldots e^{a_r}_{\mu_r}(x)
A_{a_1\ldots a_r}(x).
\end{equation}
We already know that the scalar field $A_{a_1\ldots a_r}(x)$
transforms as $D_0$ given by (\ref{eq5.5}), so
$A_{\mu_1\ldots\mu_r}(x)$ transforms in the representation
\begin{equation}
D_r = \left(D_{adj},D_{triv}\right)^{\otimes r} \otimes D_0 =
\bigoplus_j \left(D_{adj}^{\otimes r} \otimes D_j,
D_{\bar{\jmath}}\right). \label{eq5.17}
\end{equation}
Despite its asymmetric appearance the formula (\ref{eq5.17}) is really
symmetric under exchange of the two $G$ factors in $\cal G$.

{\vspace*{10mm} \large \bf \noindent 6. The spectrum of the SU(2)
string theory\\}
A conformal field theory may be interpreted as a string theory in the
following way. The physical states of the string theory are in a one
to one correspondence with the Virasoro primary states of conformal
dimension $(h,\bar{h})=(1,1)$. The amplitude for scattering of such
string states is computed by calculating the conformal field theory
correlation function of the corresponding Virasoro primary fields and
integrating it other all possible world-sheet configurations
\cite{Polyakov}. In general this implies an integration over the
insertion points on the world-sheet as well as an integration over the
space of world-sheet geometries. However, conformal invariance reduces
the latter infinite dimensional integral to a sum over the number of
handles (the genus) of the world-sheet and a finite dimensional
integral over the space of conformally inequivalent geometries (the
modulispace) for each genus. Different genera correspond to different
orders in the string coupling constant.

The algebraic structure of the $SU(2)$ Wess-Zumino-Witten conformal
field theory consists of two $SU(2)$ level $k$ Kac-Moody algebras
spanned by the current modes $J^a_n$ and $\bar{J}^a_n$, and two
enveloping Virasoro algebras spanned by $L_n$ and $\bar{L}_n$
\cite{Knizhnik-Zamolodchikov}. The commutation relations are
\begin{eqnarray}
&& [J^a_n,J^b_m] = i f^{ab}_{\;\;\;\;c} J^c_{n+m} + \frac{k}{2} n
\eta^{ab} \delta_{n+m,0} \nonumber\\ && [L_n,L_m] = (n-m)L_{n+m} +
\frac{c}{12} n (n^2 -1) \delta_{n+m,0} \label{1} \\ && [L_n,J^a_m] = -
m J^a_{n+m}, \nonumber
\end{eqnarray}
and an identical set for the $\bar{J}^a_n$ and $\bar{L}_n$ operators.
The Virasoro operators are constructed as normal ordered bilinears in
the Kac-Moody currents:
\begin{equation}
L_n =  \frac{1}{k+2} \sum_{m \in \bf Z} : J^a_{n-m} J_{a\;m} : \; .
\end{equation}
The conformal anomaly is $c=3k/(k+2)$, but, as we have already
mentioned, the value of $c$ does not concern us for the moment.

All fields in the theory could be arranged into families consisting of
an ancestor field, which is primary with respect to the extended
conformal algebra (\ref{1}), and an infinite set of descendant fields.
A primary field is characterized by the representation in which it
transforms under the isometry group ${\cal G} = SU(2) \times SU(2)$ of
the target manifold. Not all representations are allowed, however, but
only a subset of so called integrable ones \cite{Gepner-Witten}.

A representation of $SU(2)$ is labeled by a non-negative integer or
half integer spin $j$. These representations are self conjugate, and
in the following $j$ and $\bar{\jmath}$ will denote independent
representations. The representation space of $D_j$ is spanned by the
orthonormal states $|j,m>$, where $m=-j,-j+1,\ldots,j$. The tensor
product of two representations is decomposed as
\begin{equation}
D_{j_1} \otimes D_{j_2} = \bigoplus_{j=|j_1-j_2|}^{j_1+j_2} D_j.
\end{equation}
The integrable representations of the level $k$ $SU(2)$ Kac-Moody
algebra are those with $j \leq k/2$ \cite{Gepner-Witten}. They form a
closed operator product algebra \cite{Gepner-Witten} according to
\begin{equation}
D_{j_1} \times D_{j_2} = \sum_{j=|j_1-j_2|}^{\min (j_1+j_2,k-j_1-j_2)}
D_j. \label{9}
\end{equation}

The Hilbert space is built on groundstates $|j_0;m,\bar{m}>$, where
$j_o \leq k/2$, which are created by the primary fields acting on the
vacuum state $|0>$. These groundstates are annihilated by the
Kac-Moody currents with positive mode numbers, and transform in the
$(D_{j_0},D_{j_0})$ representation under the $SU(2) \times SU(2)$
generated by the zero-modes. The Hilbert space is thus spanned by
states of the form
\begin{equation}
|\Psi> = J^{a_1}_{-n_1} \ldots J^{a_s}_{-n_s}
\bar{J}^{\bar{a}_1}_{-\bar{n}_1} \ldots
\bar{J}^{\bar{a}_t}_{-\bar{n}_t} | j_0;m,\bar{m}>.
\end{equation}
The state $|\Psi>$ is an eigenstate of the $L_0$ and $\bar{L}_0$
operators with eigenvalues
\begin{equation}
h = \frac{j_0(j_0+1)}{k+2}+\sum_s n_s \;\;\;\;,\;\;\;\; \bar{h} =
\frac{j_0(j_0+1)}{k+2}+\sum_t \bar{n}_t. \label{2}
\end{equation}

It is easy to determine the spectrum of physical states, i.~e. the
Virasoro primary states of conformal dimension $(h,\bar{h})=(1,1)$,
from equation (\ref{2}). The only solutions are
\begin{equation}
|j_0;m,\bar{m}> \;\;\;\; , \;\;\;\; j_0(j_0+1) = k + 2 \label{5}
\end{equation}
and
\begin{equation}
J^a_{-1} \bar{J}^b_{-1} |0;0,0>. \label{6}
\end{equation}
We see that for a general integer level $k$, the first set of states
is absent, since the spin $j_0$ has to be integer. (Half integer spins
would require non-integer levels.) The corresponding Virasoro primary
fields are a field which is primary with respect to the left and right
Kac-Moody algebras
\begin{equation}
V^{j_0}_{m\bar{m}}(z,\bar{z}) \label{3}
\end{equation}
and a product of a holomorphic and an anti-holomorphic Kac-Moody current
\begin{equation}
V^{ab}(z,\bar{z}) = J^a(z) \bar{J}^b(\bar{z}). \label{4}
\end{equation}
The physical states (\ref{5}) and (\ref{6}) transform in the
$(j,\bar{\jmath})=(j_0,j_0)$ and $(j,\bar{\jmath})=(1,1)$
representations respectively under ${\cal G} = G \times G$.

{\vspace*{10mm} \large \bf \noindent 7. Field theory on the SU(2)
group manifold\\}
Our object in this section is to construct a local
quantum field theory on the $SU(2)$ group manifold which reproduces
the scattering amplitude of the Wess-Zumino-Witten model interpreted
as a string theory.

The first question to settle is what the field content of our theory
should be. The quantum numbers of the primary fields suggest that we
should introduce a scalar field $\sigma(x)$ and a rank two tensor
field $A_{\mu\nu}(x)$. It is convenient to decompose the tensor field
into its antisymmetric part $b_{\mu\nu}(x)$, its symmetric trace-less
part $h_{\mu\nu}(x)$ and its trace $\phi(x)$. This is in analogy with
the bosonic string in flat space, where we find a tachyon $\sigma$, an
antisymmetric tensor $b_{\mu\nu}$, a graviton $h_{\mu\nu}$ and a
dilaton $\phi$. In flat space string theory we also get an infinite
tower of massive states which have no counterparts in the $SU(2)$
string. It is easy to see that this is related to the target space
being Euclidean rather than Minkowskian in signature.

The next step is to determine the representation content of each of
these fields under the target space isometry group ${\cal G} = SU(2)
\times SU(2)$. As was explained in section~5 we rewrite
$A_{\mu\nu}(x)$ as
\begin{equation}
A_{\mu\nu}(x) = e^a_\mu(x) e^b_\nu(x) A_{ab}(x).
\end{equation}
A scalar field such as $\sigma(x)$, $\phi(x)$ or $A_{ab}(x)$
transforms under ${\cal G} = SU(2) \times SU(2)$ as
\begin{equation}
D_\sigma = D_\phi = \bigoplus_j (D_j,D_j).
\end{equation}
The vielbein $e^a_\mu(x)$ transforms as $(D_{adj},D_{triv}) =
(D_1,D_0)$. We see that the antisymmetric, symmetric traceless and
trace parts of the product of two vielbeins transform as $(D_1,D_0)$,
$(D_2,D_0)$ and $(D_0,D_0)$ respectively. The representation content
of $b_{\mu\nu}(x)$ and $h_{\mu\nu}(x)$ is thus
\begin{eqnarray}
&&D_b = \bigoplus_j (D_1 \otimes D_j , D_j ) =
\bigoplus^\prime_{|j-\bar{\jmath}| \leq 1} (D_j , D_{\bar{\jmath}} )
\nonumber\\ &&D_h = \bigoplus_j (D_2 \otimes D_j , D_j ) =
\bigoplus^\prime_{|j-\bar{\jmath}| \leq 2} (D_j , D_{\bar{\jmath}} ) .
\end{eqnarray}
(The prime on the summation symbols indicates that a few of the lowest
lying representations are missing.)

We should now find all possible kinetic terms and interaction terms
for these fields and determine how the different representations
couple together. Let us begin with the interaction terms. For the
moment we will only consider tree level contributions to scattering of
external $\sigma(x)$ fields, so we should write down possible three
point couplings between two such on-shell fields and a third arbitrary
field which may be off shell. By charge conjugation symmetry there is
no possible $b\sigma^2$ interaction. The possible $\phi \sigma^2$ and
$\sigma^3$ terms without any derivatives are just
\begin{equation}
S^{int}_{\phi\sigma\sigma} = \int d^3 x \; \lambda_\phi \phi(x) \sigma^2(x)
\end{equation}
and
\begin{equation}
S^{int}_{\sigma\sigma\sigma} = \int d^3 x \; \lambda_\sigma \sigma^3(x)
\end{equation}
respectively. The most general $h \sigma^2$ term with two derivatives is
\begin{eqnarray}
S^{int}_{h\sigma\sigma} & = & \int d^3 x \; \left( \lambda_1
h^{\mu\nu}(x) D_\mu \sigma(x) D_\nu \sigma(x) + \lambda_2
h^{\mu\nu}(x) \sigma(x) D_\mu D_\nu \sigma(x) \right) \nonumber\\ & =
& \int d^3 x \; \left( \lambda_1 h^{ab}(x) D_a \sigma(x) D_b \sigma(x)
\!+\! \lambda_2 h^{ab}(x) \sigma(x) D_a D_b \sigma(x) \right) .
\end{eqnarray}

We next turn to the kinetic terms. For $\sigma$ and $\phi$, the only
possible terms which contain at most two derivatives are
\begin{equation}
S^{kin}_{\sigma\sigma} = \int d^3 x \; \sigma(x) ( D^\mu D_\mu -
m_\sigma^2 ) \sigma(x) = \int d^3 x \; \sigma(x) (D^a D_a - m_\sigma^2
) \sigma(x)
\end{equation}
and
\begin{equation}
S^{kin}_{\phi\phi} = \int d^3 x \; \phi(x) ( D^\mu D_\mu - m_\phi^2 )
\phi(x) =\int d^3 x \; \phi(x) ( D^a D_a - m_\phi^2 ) \phi(x)
\end{equation}
respectively. The constants $m_\sigma$ and $m_\phi$ are determined by
the requirement that the propagators should diverge on shell.

For $h^{\mu\nu}$ the most general kinetic term is
\begin{eqnarray}
S^{kin}_{hh} & = & \int d^3 x \; \left( h^{\mu\nu}(x) D^\rho D_\rho
h_{\mu\nu}(x) + \kappa D_\mu h^{\mu\nu}(x) D^\rho h_{\rho\nu}(x) -
m_h^2 h^{\mu\nu}(x) h_{\mu\nu}(x) \right) \nonumber\\ & = & \int d^3 x
\; \left( h^{ab}(x) D^c D_c h_{ab}(x) + 2 i h^{ab}(x)
f^{ce}_{\;\;\;\;a} D_e h_{cb}(x) \right. \\ && + \left. \kappa D_a
h^{ab}(x) D_c h^c_{\;\;b}(x) - ( m_h^2 + \frac{3}{4} ) h^{ab}(x)
h_{ab}(x) \right) \nonumber.
\end{eqnarray}

There may also be a kinetic term of $h\phi$ type:
\begin{equation}
S^{kin}_{h\phi} = \int d^3 x \; \rho h^{\mu\nu}(x) D_\mu D_\nu \phi(x)
= \int d^3 x \; \rho h^{ab}(x) D_a D_b \phi(x).
\end{equation}

The action is invariant under ${\cal G} = SU(2) \times SU(2)$. To
evaluate the couplings between different representations in the
various terms, we need the following rules:
\begin{enumerate}
\item An integration of the product of three scalar functions yields a $3jm$
symbol for each of the two $SU(2)$ factors. We depict this as a vertex with
three incoming lines and a factor 1:
\begin{equation}
\int d^3 x \; \sigma_1(x) \sigma_2(x) \sigma_3(x) = \begin{array}{c} {\small
\begin{picture}(35,30)  \put(5,5){\line(1,1){10}} \put(5,25){\line(1,-1){10}}
\put(15,15){\line(1,0){15}} \put(0,27){$j_1$} \put(0,1){$j_2$}
\put(31,13){$j_3$}  \end{picture} } \end{array}
\end{equation}
A special case is when one of the functions is the identity 1, which
only carries the trivial representation $D_0$. The representations
from the remaining two functions then couple together to yield the
trivial representation.
\item The derivative $D_a$ acts on a scalar function as a multiplication with
the Lie algebra generator $T_a$ from the left. The matrix $T_a$ in the $D_j$
representation, in its turn, is proportional to $\sqrt{j(j+1)}$ times the
$3jm$ symbol for coupling of $D_j$ and the adjoint representation $D_1$ to
yield $D_j$. The right $SU(2)$ factor is unaffected. A graphic representation
is
\begin{equation}
D_a \sigma(x) = \begin{array}{c} {\small \begin{picture}(50,30)
\put(5,5){\line(1,1){10}} \put(5,25){\line(1,-1){10}}
\put(15,15){\line(1,0){30}} \put(0,26){$j$} \put(0,1){$j$}
\put(46,13){$1$} \put(14,21){$\sqrt{j(j+1)}$} \end{picture} }
\end{array}
\end{equation}
\item The d'Alembertian $D_a D^a$ acting on a scalar function amounts to
multiplication by the eigenvalue $j(j+1)$ of the quadratic Casimir operator.
\begin{equation}
D^a D_a \sigma(x) = \begin{array}{c} {\small \begin{picture}(45,15)
\put(5,5){\line(1,0){35}} \put(0,3){$j$} \put(41,3){$j$}
\put(10,9){$j(j+1)$} \end{picture} } \end{array}
\end{equation}
\item Contracting three Lie algebra indices with the structure constants
yields the $3jm$ symbol for coupling three adjoint representations $D_1$
together:
\begin{equation}
f^{abc} = \begin{array}{c} {\small \begin{picture}(35,30)
\put(5,5){\line(1,1){10}} \put(5,25){\line(1,-1){10}}
\put(15,15){\line(1,0){15}} \put(0,26){$1$} \put(0,1){$1$}
\put(31,13){$1$} \end{picture} } \end{array}
\end{equation}
\item A subdiagram may be rearranged by using the $6j$ symbols as in
(\ref{15}).
\end{enumerate}

The right hand $SU(2)$ factor is trivial in all the terms we have
considered. It is just a two- or three-point vertex for the kinetic
and interaction terms respectively with no additional factors.

For the left hand factors in the kinetic terms we get
\begin{eqnarray}
S^{kin}_{\phi\phi} & \sim & K_{\phi\phi}(j,j) \begin{array}{c} {\small
\begin{picture}(35,10) \put(5,5){\line(1,0){25}} \put(0,3){$j$}
\put(31,3){$j$} \end{picture} } \end{array} \nonumber\\
S^{kin}_{\sigma\sigma} & \sim & K_{\sigma\sigma}(j,j) \begin{array}{c}
{\small \begin{picture}(35,10) \put(5,5){\line(1,0){25}}
\put(0,3){$j$} \put(31,3){$j$} \end{picture} } \end{array} \nonumber\\
S^{kin}_{h\phi} & \sim & j(j+1) \begin{array}{c} {\small
\begin{picture}(45,30) \put(5,5){\line(2,1){20}}
\put(5,15){\line(2,1){10}} \put(5,25){\line(2,-1){20}}
\put(25,15){\line(1,0){15}} \put(0,1){$1$} \put(0,13){$1$}
\put(0,25){$j$} \put(19,21){$j$} \put(41,13){$j$} \end{picture} }
\end{array} \sim K_{h\phi}(j,j) \begin{array}{c} {\small
\begin{picture}(45,30) \put(5,5){\line(2,1){20}}
\put(5,15){\line(2,-1){10}} \put(5,25){\line(2,-1){20}}
\put(25,15){\line(1,0){15}} \put(0,1){$1$} \put(0,13){$1$}
\put(0,25){$j$} \put(18,6){$2$} \put(41,13){$j$} \end{picture} }
\end{array} \nonumber\\ S^{kin}_{hh} & \sim & \left(
j(j+1)-m_h^2-\frac{3}{4} \right) \begin{array}{c} {\small
\begin{picture}(60,30) \put(5,5){\line(1,0){50}}
\put(5,15){\line(1,0){50}} \put(5,25){\line(1,0){50}} \put(0,3){$1$}
\put(0,13){$1$} \put(0,23){$j$} \put(56,3){$1$} \put(56,13){$1$}
\put(56,23){$j$} \end{picture} } \end{array}\\ && + 2 i \sqrt{j(j+1)}
\begin{array}{c} {\small \begin{picture}(60,30)
\put(5,5){\line(1,0){50}} \put(5,15){\line(1,0){50}}
\put(5,25){\line(1,0){50}} \put(30,15){\line(0,1){10}} \put(0,3){$1$}
\put(0,13){$1$} \put(0,23){$j$} \put(56,3){$1$} \put(56,13){$1$}
\put(56,23){$j$} \put(32,17){$1$} \end{picture} } \end{array} + \kappa
j(j+1) \begin{array}{c} {\small \begin{picture}(60,30)
\put(5,5){\line(1,0){50}} \put(5,15){\line(2,1){10}}
\put(5,25){\line(2,-1){10}} \put(15,20){\line(1,0){30}}
\put(45,20){\line(2,1){10}} \put(45,20){\line(2,-1){10}}
\put(0,3){$1$} \put(0,13){$1$} \put(0,24){$j$} \put(56,3){$1$}
\put(56,13){$1$} \put(56,24){$j$} \put(28,14){$j$} \end{picture} }
\end{array} \nonumber\\ & \sim & \sum_{\bar{\jmath}=j-2}^{j+2}
K_{hh}(j,\bar{\jmath}) \begin{array}{c} {\small \begin{picture}(60,30)
\put(5,5){\line(2,1){20}} \put(5,15){\line(2,-1){10}}
\put(5,25){\line(2,-1){20}} \put(25,15){\line(1,0){10}}
\put(35,15){\line(2,1){20}} \put(35,15){\line(2,-1){20}}
\put(45,10){\line(2,1){10}} \put(0,3){$1$} \put(0,13){$1$}
\put(0,24){$j$} \put(56,3){$1$} \put(56,13){$1$} \put(56,24){$j$}
\put(28,18){$\bar{\jmath}$} \put(19,7){$2$} \put(35,7){$2$}
\end{picture} } \end{array}, \nonumber
\end{eqnarray}
where the inverse propagators are given by \newpage
\begin{eqnarray}
K_{\sigma\sigma}(j,j) & = & j(j+1)-m_\sigma^2 \nonumber\\
K_{\phi\phi}(j,j) & = & j(j+1)-m_\phi^2 \nonumber\\ K_{h\phi}(j,j) & =
& \rho \left( 3 - 4j(j+1) \right) \sqrt{\frac{j(j+1)}{(2j+3)(2j-1)}}
\\ K_{hh}(j,\bar{\jmath}) & = & \sqrt{\frac{2\bar{\jmath}+1}{2j+1}}
\left( j(j+1)+\bar{\jmath}(\bar{\jmath}+1)-2m_h^2-\frac{15}{2} \right.
\nonumber\\ && \left. + \kappa
(j\!+\!\bar{\jmath}\!+\!3)(j\!+\!\bar{\jmath}\!-\!1)
(j\!-\!\bar{\jmath}\!+\!2)(\bar{\jmath}\!-\!j\!+\!2)
\begin{array}{c} \\ \\ \end{array} \right) . \nonumber
\end{eqnarray}

For the interaction terms, we are only interested in the case where
the external $\sigma(x)$ fields are on shell, i.~e. we need only
determine the couplings for their $(D_{j_0},D_{j_0})$ components,
where $j_0(j_0+1) = k+2$. Furthermore, the $j_0$ (or equivalently the
$k$) dependence may be absorbed in a renormalization of the different
coupling constants. We thus get
\begin{eqnarray}
S^{int}_{\sigma\sigma\sigma} & \sim &
V^{\sigma\sigma\sigma}(j,\bar{\jmath}) \begin{array}{c} {\small
\begin{picture}(35,30) \put(5,5){\line(1,1){10}}
\put(5,25){\line(1,-1){10}} \put(15,15){\line(1,0){15}}
\put(0,27){$j_0$} \put(0,1){$j_0$} \put(31,13){$\bar{\jmath}$}
\end{picture} } \end{array} \nonumber\\ S^{int}_{\phi\sigma\sigma} &
\sim & V^{\phi\sigma\sigma}(j,\bar{\jmath}) \begin{array}{c} {\small
\begin{picture}(35,30) \put(5,5){\line(1,1){10}}
\put(5,25){\line(1,-1){10}} \put(15,15){\line(1,0){15}}
\put(0,27){$j_0$} \put(0,1){$j_0$} \put(31,13){$\bar{\jmath}$}
\end{picture} } \end{array} \\ S^{int}_{h\sigma\sigma} & \sim &
\lambda_1 \begin{array}{c} {\small \begin{picture}(50,30)
\put(5,5){\line(1,0){20}} \put(5,15){\line(2,1){20}}
\put(5,25){\line(2,-1){8}} \put(17,19){\line(2,-1){8}}
\put(25,5){\line(0,1){20}} \put(25,5){\line(4,1){20}}
\put(25,25){\line(4,-1){20}} \put(0,3){$1$} \put(0,13){$1$}
\put(0,23){$j$} \put(46,8){$j_0$} \put(46,18){$j_0$} \put(26,9){$j_0$}
\put(26,18){$j_0$} \end{picture} } \end{array} + \lambda_2
\begin{array}{c} {\small \begin{picture}(50,30)
\put(5,5){\line(1,0){20}} \put(5,15){\line(1,0){20}}
\put(5,25){\line(1,0){20}} \put(25,5){\line(0,1){20}}
\put(25,5){\line(4,1){20}} \put(25,25){\line(4,-1){20}} \put(0,3){$1$}
\put(0,13){$1$} \put(0,23){$j$} \put(46,8){$j_0$} \put(46,18){$j_0$}
\put(26,9){$j_0$} \put(26,18){$j_0$} \end{picture} } \end{array}
\nonumber\\ & \sim & \sum_{\bar{\jmath}=j-2}^{j+2}
V^{h\sigma\sigma}(j,\bar{\jmath}) \begin{array}{c} {\small
\begin{picture}(50,30) \put(5,5){\line(2,1){20}}
\put(5,15){\line(2,-1){10}} \put(5,25){\line(2,-1){20}}
\put(25,15){\line(1,0){10}} \put(35,15){\line(2,1){10}}
\put(35,15){\line(2,-1){10}} \put(0,3){$1$} \put(0,13){$1$}
\put(0,23){$j$} \put(46,8){$j_0$} \put(46,18){$j_0$} \put(19,6){$2$}
\put(27,10){$\bar{\jmath}$} \end{picture} } \end{array}, \nonumber
\end{eqnarray}
where the vertex factors are given by
\begin{eqnarray}
V^{\sigma\sigma\sigma}(j,\bar{\jmath}) & = & \delta_{j,\bar{\jmath}}
\lambda_\sigma \nonumber\\ V^{\phi\sigma\sigma}(j,\bar{\jmath}) & = &
\delta_{j,\bar{\jmath}} \lambda_\phi \nonumber\\
V^{h\sigma\sigma}(j,\bar{\jmath}) & = & \delta_{j,\bar{\jmath}}
\sqrt{\frac{(2j+2)(2j)}{(2j\!+\!3)(2j\!-\!1)}} \left(
(3\lambda_2-\lambda_1)j(j+1)-4(\lambda_1+\lambda_2)j_0(j_0+1)-3\lambda_2
\right) \nonumber\\ && + \delta_{j-1,\bar{\jmath}}
\sqrt{\frac{3((2j)(2j_0-j+1)(2j_0+j+1)}{2(2j+2)(2j+1)(2j-2)}}
\lambda_2 \nonumber\\ && - \delta_{j+1,\bar{\jmath}}
\sqrt{\frac{3((2j+2)(2j_0-j)(2j_0+j+2)}{2(2j+4)(2j+1)(2j)}} \lambda_2
\\ && + \delta_{j-2,\bar{\jmath}} \sqrt{\frac{6 j
(j\!-\!1)(2j_0\!-\!j\!+\!2)(2j_0\!-\!j\!+\!1)(2j_0\!+\!j\!+\!1)(2j_0\!+\!j)}
{(2j\!+\!1)(2j\!-\!1)}}
(\lambda_1\!+\!\lambda_2) \nonumber\\ && + \delta_{j+2,\bar{\jmath}}
\sqrt{\frac{6
(j\!+\!2)(j\!+\!1)(2j_0\!-\!j)(2j_0\!-\!j\!-\!1)(2j_0\!+\!j\!+\!3)
(2j_0\!+\!j\!+\!2)}{(2j\!+\!3)(2j\!+\!1)}} \nonumber\\ &&
(\lambda_1+\lambda_2). \nonumber
\end{eqnarray}

The different coupling constants have been renormalized as compared to
their original definitions in the action terms. The asymmetry between
the left and the right $SU(2)$ factor is due to our normalization
conventions and will disappear when we put the propagators and the
vertices together. We get the general form of the $s$-channel
contribution to four-$\sigma$ scattering at tree level as
\begin{eqnarray}
\Gamma^{(s)}(j,\bar{\jmath}) & = & V^{\sigma\sigma\sigma}(j,\bar{\jmath})
K^{-1}_{\sigma\sigma}(j,\bar{\jmath}) V^{\sigma\sigma\sigma}(j,\bar{\jmath})
+ V^{\phi\sigma\sigma}(j,\bar{\jmath}) K^{-1}_{\phi\phi}(j,\bar{\jmath})
V^{\phi\sigma\sigma}(j,\bar{\jmath}) \\
&& + V^{h\sigma\sigma}(j,\bar{\jmath}) K^{-1}_{hh}(j,\bar{\jmath})
V^{h\sigma\sigma}(j,\bar{\jmath}) + V^{h\sigma\sigma}(j,\bar{\jmath})
K^{-1}_{h\phi}(j,\bar{\jmath}) V^{\phi\sigma\sigma}(j,\bar{\jmath}).
\nonumber
\end{eqnarray}
As we have previously explained, we should add the $t$- and
$u$-channel contributions.

{\vspace*{10mm} \large \bf \noindent 8. Calculation of tensor particle
scattering amplitudes\\} We should now compute the amplitudes for
scattering of the physical string states we found in section 3, and
compare them to the field theory amplitudes of the previous section.

To each external physical state corresponds a dimension $(h,\bar{h}) =
(1,1)$ vertex operator, as we have already described. To calculate a
string scattering amplitude we integrate the conformal field theory
correlation function of the corresponding vertex operators over their
insertion points on the world-sheet. The infinite volume of the global
conformal group will lead to a divergence, however, which we remove by
fixing three of the insertion points. If the genus of the world-sheet
is greater than zero, we should also integrate over the appropriate
moduli space of conformally inequivalent Riemann surfaces. For genus
zero (tree level) we get the $n$-point amplitude as
\begin{eqnarray}
\Gamma & = & \int \prod_{i=1}^n d^2 z_i \; \delta^2(z_A-z_A^0)
\delta^2(z_B-z_B^0) \delta^2(z_B-z_B^0) \label{101}\\
&& |(z_A-z_B)(z_B-z_C)(z_C-z_A)|^2 <V_1(z_1,\bar{z}_1) \ldots
V_n(z_n,\bar{z}_n)>. \nonumber
\end{eqnarray}
The extra factor is a Jacobian, which arises upon fixing the values of
$z_A$, $z_B$ and $z_C$. Note that the conformal dimension of $d^2 z_i$
is $(h,\bar{h}) = (-1,-1)$, so the amplitude is conformally invariant.

Our vertex operators are the Kac-Moody primary fields (\ref{3}) and
the Kac-Moody current bilinears (\ref{4}). Correlation functions
involving the latter fields are related to correlation functions of
Kac-Moody primaries by the current algebra Ward identity
\cite{Knizhnik-Zamolodchikov}
\begin{equation}
< J^a(z) V_1(z_1,\bar{z}_1) \ldots V_n(z_n,\bar{z}_n) > = \sum_{i=1}^n
\frac{T^a_i}{z-z_i} < V_1(z_1,\bar{z}_1) \ldots V_n(z_n,\bar{z}_n) >
\end{equation}
and its anti-holomorphic counterpart. As a simple example of how this
works, we consider scattering of three scalar particles and one tensor
particle. According to our previous discussions, such an amplitude is
given by
\begin{eqnarray}
\Gamma^{ab}_{m_i\bar{m}_i} & = & |(z_1-z_2)(z_2-z_3)(z_3-z_1)|^2 \nonumber\\
&& \int d^2 z \; < J^a(z)\bar{J}^b(\bar{z})
V_{m_1\bar{m}_1}(z_1,\bar{z}_1) V_{m_2\bar{m}_2}(z_2,\bar{z}_2)
V_{m_3\bar{m}_3}(z_3,\bar{z}_3) > \nonumber\\ & = &
|(z_1-z_2)(z_2-z_3)(z_3-z_1)|^2 \int d^2 z \;
\sum_{i,\bar{\imath}=1}^3 T^a_i T^b_{\bar{\imath}} (z-z_i)^{-1}
(\bar{z}-\bar{z}_i)^{-1} \\ && < V_{m_1\bar{m}_1}(z_1,\bar{z}_1)
V_{m_2\bar{m}_2}(z_2,\bar{z}_2) V_{m_3\bar{m}_3}(z_3,\bar{z}_3) >
\nonumber\\ & = & \int d^2 z \; \sum_{i,\bar{\imath}=1}^3 T^a_i
T^b_{\bar{\imath}} (z-z_i)^{-1} (\bar{z}-\bar{z}_i)^{-1} \left(
\begin{array}{ccc} j_0 & j_0 & j_0 \\ m_1 & m_2 & m_3 \end{array}
\right) \left( \begin{array}{ccc} j_0 & j_0 & j_0 \\ \bar{m}_1 &
\bar{m}_2 & \bar{m}_3 \end{array} \right), \nonumber
\end{eqnarray}
where we have used the $3jm$ symbols in the last step. We now put
$(z_1,z_2,z_3) = (0,1,\infty)$, and use the property that the $T^a$
and $T^b$ matrices are the Clebsch-Gordan coefficients for coupling
spin $j_0$ and spin $1$ to yield spin $j_0$ for the left and right
$SU(2)$ factor respectively. Our amplitude could then symbolically be
written as
\begin{equation}
\Gamma = \int d^2 z  \left[ z^{-1} \!\! \begin{array}{c} {\small
\begin{picture}(30,30)  \put(5,5){\line(1,2){5}} \put(5,25){\line(1,-2){5}}
\put(20,15){\line(1,-2){5}} \put(20,15){\line(1,2){5}}
\put(10,15){\line(1,0){10}} \put(0,26){$1$} \put(26,26){$j_0$}
\put(0,1){$j_0$} \put(26,1){$j_0$}  \put(13,9){$j_0$} \end{picture} }
\end{array} \!\!+\! (z-1)^{-1} \!\! \begin{array}{c} {\small
\begin{picture}(30,30)  \put(5,5){\line(2,1){10}} \put(5,25){\line(2,-1){10}}
\put(15,10){\line(2,-1){10}} \put(15,20){\line(2,1){10}}
\put(15,10){\line(0,1){10}}  \put(0,26){$1$} \put(26,26){$j_0$}
\put(0,1){$j_0$} \put(26,1){$j_0$} \put(8,13){$j_0$} \end{picture} }
\end{array} \right] \times \left[ \bar{z}^{-1} \!\! \begin{array}{c} {\small
\begin{picture}(30,30)  \put(5,5){\line(1,2){5}} \put(5,25){\line(1,-2){5}}
\put(20,15){\line(1,-2){5}} \put(20,15){\line(1,2){5}}
\put(10,15){\line(1,0){10}} \put(0,26){$1$} \put(26,26){$j_0$}
\put(0,1){$j_0$} \put(26,1){$j_0$}  \put(13,9){$j_0$} \end{picture} }
\end{array} \!\!+\! (\bar{z}-1)^{-1} \!\! \begin{array}{c} {\small
\begin{picture}(30,30)  \put(5,5){\line(2,1){10}} \put(5,25){\line(2,-1){10}}
\put(15,10){\line(2,-1){10}} \put(15,20){\line(2,1){10}}
\put(15,10){\line(0,1){10}}  \put(0,26){$1$} \put(26,26){$j_0$}
\put(0,1){$j_0$} \put(26,1){$j_0$} \put(8,13){$j_0$} \end{picture} }
\end{array} \right],
\end{equation}
where the first and second bracket corresponds to the left and right
$SU(2)$ factors respectively.

The regularization of integrals like those above will be discussed in
section~10. We require that
\begin{eqnarray}
&& \int d^2 z \; z^{-1} \bar{z}^{-1} = \int d^2 z \; (z\!-\!1)^{-1}
(\bar{z}\!-\!1)^{-1} \\ && \hspace*{10mm} = 2 \int d^2 z \; z^{-1}
(\bar{z}\!-\!1)^{-1} = 2 \int d^2 z \; (z\!-\!1)^{-1} \bar{z}^{-1} = 2
\delta, \nonumber
\end{eqnarray}
where the constant $\delta$ diverges as the regulator is turned off.
We thus get
\begin{eqnarray}
\Gamma & \sim & \begin{array}{c} {\small \begin{picture}(30,30)
\put(5,5){\line(1,2){5}} \put(5,25){\line(1,-2){5}}
\put(20,15){\line(1,-2){5}} \put(20,15){\line(1,2){5}}
\put(10,15){\line(1,0){10}} \put(0,26){$1$} \put(26,26){$j_0$}
\put(0,1){$j_0$} \put(26,1){$j_0$}  \put(13,9){$j_0$} \end{picture} }
\end{array} \!\times\! \begin{array}{c} {\small \begin{picture}(30,30)
\put(5,5){\line(1,2){5}} \put(5,25){\line(1,-2){5}}
\put(20,15){\line(1,-2){5}} \put(20,15){\line(1,2){5}}
\put(10,15){\line(1,0){10}} \put(0,26){$1$} \put(26,26){$j_0$}
\put(0,1){$j_0$} \put(26,1){$j_0$}  \put(13,9){$j_0$} \end{picture} }
\end{array} \;+\; \begin{array}{ccc} {\small \begin{picture}(30,30)
\put(5,5){\line(2,1){10}} \put(5,25){\line(2,-1){10}}
\put(15,10){\line(2,-1){10}} \put(15,20){\line(2,1){10}}
\put(15,10){\line(0,1){10}}  \put(0,26){$1$} \put(26,26){$j_0$}
\put(0,1){$j_0$} \put(26,1){$j_0$} \put(8,13){$j_0$} \end{picture} }
\end{array} \!\times\! \begin{array}{c} {\small \begin{picture}(30,30)
\put(5,5){\line(2,1){10}} \put(5,25){\line(2,-1){10}}
\put(15,10){\line(2,-1){10}} \put(15,20){\line(2,1){10}}
\put(15,10){\line(0,1){10}}  \put(0,26){$1$} \put(26,26){$j_0$}
\put(0,1){$j_0$} \put(26,1){$j_0$} \put(8,13){$j_0$} \end{picture} }
\end{array} \\
&& +\frac{1}{2} \begin{array}{c} {\small \begin{picture}(30,30)
\put(5,5){\line(1,2){5}} \put(5,25){\line(1,-2){5}}
\put(20,15){\line(1,-2){5}} \put(20,15){\line(1,2){5}}
\put(10,15){\line(1,0){10}} \put(0,26){$1$} \put(26,26){$j_0$}
\put(0,1){$j_0$} \put(26,1){$j_0$} \put(13,9){$j_0$} \end{picture} }
\end{array} \!\times\! \begin{array}{c} {\small \begin{picture}(30,30)
\put(5,5){\line(2,1){10}} \put(5,25){\line(2,-1){10}}
\put(15,10){\line(2,-1){10}} \put(15,20){\line(2,1){10}}
\put(15,10){\line(0,1){10}} \put(0,26){$1$} \put(26,26){$j_0$}
\put(0,1){$j_0$} \put(26,1){$j_0$} \put(8,13){$j_0$} \end{picture} }
\end{array} \;+\; \frac{1}{2} \begin{array}{c} {\small
\begin{picture}(30,30) \put(5,5){\line(2,1){10}}
\put(5,25){\line(2,-1){10}} \put(15,10){\line(2,-1){10}}
\put(15,20){\line(2,1){10}} \put(15,10){\line(0,1){10}}
\put(0,26){$1$} \put(26,26){$j_0$} \put(0,1){$j_0$} \put(26,1){$j_0$}
\put(8,13){$j_0$} \end{picture} } \end{array} \!\times\!
\begin{array}{c} {\small \begin{picture}(30,30)
\put(5,5){\line(1,2){5}} \put(5,25){\line(1,-2){5}}
\put(20,15){\line(1,-2){5}} \put(20,15){\line(1,2){5}}
\put(10,15){\line(1,0){10}} \put(0,26){$1$} \put(26,26){$j_0$}
\put(0,1){$j_0$} \put(26,1){$j_0$} \put(13,9){$j_0$} \end{picture} }
\end{array} \nonumber.
\end{eqnarray}
An evaluation of the $6j$ symbols for the coupling of three spin $j_0$
and one spin $1$ yields (see equation \ref{15})
\begin{equation}
\begin{array}{c} {\small \begin{picture}(30,30)  \put(5,5){\line(1,2){5}}
\put(5,25){\line(1,-2){5}} \put(20,15){\line(1,-2){5}}
\put(20,15){\line(1,2){5}} \put(10,15){\line(1,0){10}} \put(0,27){$1$}
\put(26,27){$j_0$} \put(0,1){$j_0$} \put(26,1){$j_0$}  \put(13,9){$j$}
\end{picture} } \end{array} = \sum_{j^\prime=j_0-1}^{j_0+1} c_{j,j^\prime}
\begin{array}{c} {\small \begin{picture}(30,30)  \put(5,5){\line(2,1){10}}
\put(5,25){\line(2,-1){10}} \put(15,10){\line(2,-1){10}}
\put(15,20){\line(2,1){10}} \put(15,10){\line(0,1){10}}  \put(0,27){$1$}
\put(26,27){$j_0$} \put(0,1){$j_0$} \put(26,1){$j_0$} \put(9,12){$j^\prime$}
\end{picture} } \end{array}
\end{equation}
where
\begin{eqnarray}
&&\left( \begin{array}{ccc}
c_{j_0-1,j_0-1} & c_{j_0-1,j_0} & c_{j_0-1,j_0+1}\\ c_{j_0,j_0-1} &
c_{j_0,j_0} & c_{j_0,j_0+1}\\ c_{j_0+1,j_0-1} & c_{j_0+1,j_0} &
c_{j_0+1,j_0+1}
\end{array} \right) \\ && = \frac{(-1)^{3j_0+1}}{2(2j_0\!+\!1)} \left(
\begin{array}{ccc} j_0+1 & \sqrt{(3j_0\!+\!1)(2j_0\!+\!1)} &
\sqrt{(3j_0\!+\!2)(3j_0\!+\!1)} \\ \sqrt{(3j_0\!+\!1)(2j_0\!+\!1)} &
(2j_0\!+\!1) & - \sqrt{(3j_0\!+\!2)(2j_0\!+\!1)} \\
\sqrt{(3j_0\!+\!2)(3j_0\!+\!1)} & - \sqrt{(3j_0\!+\!2)(2j_0\!+\!1)} &
j_0 \end{array} \right). \nonumber
\end{eqnarray}
If we also introduce
\begin{equation}
\left( \begin{array}{ccc} d_{j_0-1,j_0-1}&d_{j_0-1,j_0}&d_{j_0-1,j_0+1}\\
d_{j_0,j_0-1}&d_{j_0,j_0}&d_{j_0,j_0+1}\\d_{j_0+1,j_0-1}&d_{j_0+1,j_0} &
d_{j_0+1,j_0+1} \end{array} \right) = (-1)^{3j_0+1} \left( \begin{array}{ccc}
1&0&0\\ 0&-1&0\\0&0&1 \end{array} \right),
\end{equation}
we may express our scattering amplitude in the $s$-channel basis as
\begin{equation}
\Gamma_{j\bar{\jmath}} \sim c_{j_0,j} c_{j_0,\bar{\jmath}} + d_{j_0,j}
d_{j_0,\bar{\jmath}} + \frac{1}{2} c_{j_0,j} d_{j_0,\bar{\jmath}} +
\frac{1}{2} d_{j_0,j} c_{j_0,\bar{\jmath}},
\end{equation}
or in matrix form
\begin{equation}
\Gamma \sim \left( \begin{array}{ccc} 3j_0+1 & 0 & -\sqrt{(3j_0+1)(3j_0+2)} \\
0 & 3(2j_0+1) & 0 \\ -\sqrt{(3j_0+1)(3j_0+2)} & 0 & 3j_0+2 \end{array}
\right). \label{10}
\end{equation}
In a quantum field theory we expect a pole at $j=\bar{\jmath}=j_0$ in
each channel due to exchange of a scalar field quanta. The divergent
part of the $s$-channel contribution should therefore be proportional
to
\begin{equation}
\Gamma^{(s)} = \left( \begin{array}{ccc} 0&0&0\\0&1&0\\0&0&0 \end{array}
\right).
\end{equation}
To include the contributions from the other channels, we should
symmetrize the $s$-channel contribution with respect to the
permutation group of the external legs. The total amplitude is thus
\begin{equation}
\Gamma \sim \sum_{p \in P} p \Gamma^{(s)} p^{\rm T}, \label{11}
\end{equation}
where $P$ is the matrix group generated by the matrices $c$ and $d$.
An explicit evaluation of (\ref{11}) shows that this amplitude is in
fact proportional to (\ref{10}). The field theoretical amplitude
should also include finite contributions when the exchanged quanta is
not on shell. Since our string theory amplitude only contains
divergent parts, we must conclude that the finite parts from exchange
of scalar and tensor particles cancel. Observe that the kinematics are
such, that the latter particles cannot be on shell, so we get no
further infinite contributions.

{\vspace*{10mm} \large \bf \noindent 9. Calculation of scalar particle
correlation functions\\} Correlation functions of Kac-Moody primary
fields are in principle determined by the Knizhnik-Zamolodchikov
equations
\begin{equation}
\left( \frac{\partial}{\partial z_i} - \frac{2}{k+2} \sum_{j \neq i} T^a_i
T_{aj} (z_i-z_j)^{-1} \right) < V_1(z_1,\bar{z}_1) \ldots V_n(z_n,\bar{z}_n) >
\end{equation}
and their anti-holomorphic counterparts when supplemented by the
requirement that the correlation functions be well-defined on the
punctured Riemann sphere.

We will only consider four-point functions, where these equations
amount to ordinary differential equations, since global conformal
invariance allows us to fix three of the insertion points
\cite{Koba-Nielsen}\cite{Belavin-Polyakov-Zamolodchikov}. Indeed, a
four-point function takes the form
\begin{equation}
< V_1(z_1,\bar{z}_1) \ldots V_4(z_4,\bar{z}_4) > = |z_1-z_4|^{-4}
|z_2-z_3|^{-4} f(\eta,\bar{\eta}). \label{eq7.5}
\end{equation}
Here the crossratio $\eta$ is defined as
\begin{equation}
\eta = (z_1-z_2)(z_3-z_4)(z_3-z_2)^{-1}(z_1-z_4)^{-1}, \label{eq7.6}
\end{equation}
and the functions $f(\eta,\bar{\eta})$ fulfil
\begin{equation}
\frac{\partial}{\partial\eta}f(\eta,\bar{\eta})= \left( \frac{A}{\eta} +
\frac{B}{1-\eta} \right) f(\eta,\bar{\eta}), \label{7}
\end{equation}
where $A=-2(k+2)^{-1}T^a_1T_{a2}$ and $B=2(k+2)^{-1}T^a_1T_{a3}$, and
a similar equation for the $\bar{\eta}$-dependence. This is a linear
matrix differential equation with (regular) singular points in $\eta =
0,1,\infty$. (See for example \cite{Hille}.)

The functions $f(\eta,\bar{\eta})$ transform in the $D_{j_0} \otimes
\ldots \otimes D_{j_0}$ representations under the left and right
$SU(2)$ groups. The invariance of the correlation function tells us
that the components of $f(\eta,\bar{\eta})$ that transform
non-trivially vanish. The remaining components may conveniently be
labeled by the spins $j$ and $\bar{\jmath}$, ranging from $0$ to
$2j_0$, of the left and right $SU(2)$ representations which are
exchanged in one the ``channels'', for example $1,2\rightarrow3,4$. In
this basis the $A$ matrix is diagonal with elements
\begin{equation}
(A)_{j,j} = \frac{j(j+1)}{j_0(j_0+1)}-2
\end{equation}
and $B$ is tri-diagonal \cite{Christe-Flume} with diagonal elements
\begin{equation}
(B)_{j,j} = \frac{j(j+1)}{2j_0(j_0+1)}
\end{equation}
and off-diagonal elements
\begin{equation}
(B)_{j-1,j}=(B)_{j,j-1} = \frac{j(j^2-(2j_0+1)^2)}{2j_0(j_0+1)\sqrt{4j^2-1}}.
\end{equation}

If we introduce fundamental solutions $X^{(i)}$ to (\ref{7}) of the form
\begin{equation}
X^{(i)}_j(\eta) = \eta^{(A)_{i,i}}(\delta^i_j+{\cal O}(\eta^{|i-j|})).
\label{8}
\end{equation}
we may write the functions $f_{j\bar{\jmath}}$ as
\begin{equation}
f_{j\bar{\jmath}}(\eta,\bar{\eta}) = \sum_{i,\bar{\imath}}
c_{i\bar{\imath}} X^{(i)}_j(\eta)
X^{(\bar{\imath})}_{\bar{\jmath}}(\bar{\eta}).
\end{equation}

The coefficients $c_{i\bar{\imath}}$ are determined by the requirement
that the correlation functions be well-defined when we impose that
$\bar{\eta}$ is the complex conjugate of $\eta$. To get a well-defined
correlation function as $\eta \rightarrow 0$ we choose the matrix
$c_{i\bar{\imath}}$ diagonal, i.~e.
$c_{i\bar{\imath}}=\delta_{i,\bar{\imath}}c_i$. (There might also be
non-diagonal solutions but they need not concern us here.) If one of
the fundamental solutions $X^{(i)}(\eta)$ is continued analytically
around one of the singular points $\eta=0,1,\infty$ it is transformed
into a linear combination of solutions:
\begin{equation}
X^{(i)}(\eta) \mapsto \sum_n \alpha_{in} X^{(n)}(\eta). \label{17a}
\end{equation}
The coefficients $c_i$ should now be chosen so that the diagonal form
of $c_{i\bar{\imath}}$ is preserved, i.~e. we must require that
\begin{equation}
\sum_k c_k \alpha_{kn} \alpha_{km} = 0 \;\;\;,\;\;\; n \neq m. \label{17}
\end{equation}
This is quite different from string theory in flat space, where the
correlation functions are simply a product of a holomorphic and an
anti-holomorphic factor.

It is easy to see, that for analytic continuation around $\eta=0$, the
monodromy matrix $\alpha_{i\bar{\imath}}$ is diagonal with elements
$\alpha_{ii} = \exp(2\pi i (A)_{i,i})$. For analytic continuation
around $\eta=1$ or $\eta=\infty$ the problem is more difficult, but
nevertheless tractable. The fundamental solutions (\ref{8}) may be
expressed in terms of multiple integrals of Euler type
\cite{Christe-Flume} , which are suitable for analytic continuation
\cite{Dotsenko-Fateev}, and the coefficients $\alpha_{i\bar{\imath}}$
may be calculated.

The correlation functions may thus be expressed in terms of multiple
($2j_0$-tuple) integrals
\cite{Christe-Flume}\cite{Zamolodchikov-Fateev}. This form is not well
suited for integrating over the insertion points on the world-sheet,
however. We have not found any way to perform such integrals
analytically, and probably one has to resort to numerical methods. We
expect to come back to this issue shortly.

{\vspace*{10mm} \large \bf \noindent 10.  The k=4 case\\}
The computation of the $c_{i\bar{\imath}}$ coefficients is greatly
facilitated in the case that $k-j_3-j_4=j_4-j_3$, so that the operator
product expansion of primary fields with spins $j_3$ and $j_4$ only
contains fields which are descendant from spin $j_4-j_3$. This means
that $c_{i\bar{\imath}}=0$ except for $i=\bar{\imath}=j_4-j_3$.

In particular, if $j_1=j_2=j_3=j_4=j_0$ and $k=2j_0$, we have
\begin{eqnarray}
(A)_{jj} & = & \frac{j(j+1)}{2(j_0+1)}-j_0 \;\;,\;\; j =0,1,\ldots,2j_0 \\
(B)_{jj} & = & \frac{j(j+1)}{4(j_0+1)} \nonumber
\end{eqnarray}
and
\begin{equation}
(B)_{j,j-1}=(B)_{j-1,j}=\frac{j((2j_0+1)^2-j^2)}{4(j_0+1)\sqrt{4j^2-1}}
\;\;,\;\; j=1,2,\ldots,2j_0.
\end{equation}
The relevant solution to the Knizhnik-Zamolodchikov equation is
\begin{equation}
X_j(\eta)=\sqrt{2j+1}\frac{(2j_0-j)!}{j!(2j_0)!} \sum_{s=0}^{2j_0-j}
\frac{(j+s)!(2j_0-s)!}{(2j_0-j-s)!s!} \eta^{j-j_0}(1-\eta)^{s-j_0}.
\label{eq7.13}
\end{equation}

In string theory we need $h=1$ for our primary fields, so according to
equation (\ref{2}) we must take $k=j_0(j_0+1)-2$. To be able to use
the simplified solution procedure described in the last paragraph we
must also have $k=2j_0$, which yields $j_0=2$ and $k=4$. The solution
(\ref{eq7.13}) is in this case
\begin{eqnarray}
X_0(\eta) = \sqrt{1} & \eta^{-2} & (
(1-\eta)^{-2}+(1-\eta)^{-1}+1+(1-\eta)+ (1-\eta)^2) \nonumber\\
X_1(\eta) = \sqrt{3} & \eta^{-1} & ( (1-\eta)^{-2}+\frac{3}{2}
(1-\eta)^{-1} +\frac{3}{2} + (1-\eta)) \nonumber\\ X_2(\eta) =
\sqrt{5} & \eta^0& ( (1-\eta)^{-2} + \frac{3}{2} (1-\eta)^{-1} + 1) \\
X_3(\eta) = \sqrt{7} & \eta^1 & ( (1-\eta)^{-2} + (1-\eta)^{-1})
\nonumber\\ X_4(\eta) = \sqrt{9} & \eta^2 & (1-\eta)^{-2}. \nonumber
\end{eqnarray}
It will prove convenient to perform a partial fraction decomposition
of these functions and write them as
\begin{eqnarray}
X_0(\eta) & = & 1 + 5 \eta^{-2} + 3 (1-\eta)^{-1} + (1-\eta)^{-2}
\nonumber\\ X_1(\eta) & = &
\sqrt{3}(-1+5\eta^{-1}+\frac{5}{2}(1-\eta)^{-1}+(1-\eta)^{-2})
\nonumber\\ X_2(\eta) & = &
\sqrt{5}(1+\frac{3}{2}(1-\eta)^{-1}+(1-\eta)^{-2})\\ X_3(\eta) & = &
\sqrt{7}(-1+(1-\eta)^{-2}) \nonumber\\ X_4(\eta) & = &
3(1-2(1-\eta)^{-1}+(1-\eta)^{-2})\nonumber.
\end{eqnarray}
The complete correlation function follows from (\ref{eq7.5}):
\begin{eqnarray}
&&<V_{m_1\bar{m}_1}(z_1,\bar{z}_1)\ldots
V_{m_4\bar{m}_4}(z_4,\bar{z}_4)> \label{eq7.16}\\ &&\hspace*{10mm} =
\sum_{j\bar{\jmath}} P^j_{m_1\ldots m_4} P^{\bar{\jmath}}_{\bar{m}_1
\ldots \bar{m}_4} |(z_1-z_4)(z_2-z_3)|^{-4} X_j(\eta)
X_{\bar{\jmath}}(\bar{\eta}), \nonumber
\end{eqnarray}
where $P^j_{m_1\ldots m_4}$ projects on the invariant in
$(D_2)^{\otimes 4}$ which has $D_j$ as intermediate representation in
the $s$-channel, and $\eta$ is given by (\ref{eq7.6}).

Integrating the correlation function (\ref{eq7.16}) with the measure
in equation (\ref{101}), choosing $(z_1^0,z_2^0,z_4^0)=(\infty,1,0)$
and changing the integration variable to $\eta$ as defined in
(\ref{eq7.6}), we get the amplitude
\begin{equation}
\Gamma_{m_1 \ldots m_4 \bar{m}_1 \ldots \bar{m}_4} = \sum_{j\bar{\jmath}}
P^j_{m_1 \ldots m_4} P^{\bar{\jmath}}_{\bar{m}_1 \ldots \bar{m}_4}
\Gamma_{j\bar{\jmath}},
\end{equation}
where the matrix $\Gamma_{j\bar{\jmath}}$, is given by
\begin{equation}
\Gamma_{j\bar{\jmath}} =\int \frac{d^2 \eta}{\pi} \; X_j(\eta)X_{\bar{\jmath}}
(\bar{\eta}). \label{eq8.3}
\end{equation}
We note that, formally, this amplitude is invariant under permutation
of the $s$-, $t$- and $u$-channels. Namely, such transformations
correspond to replacing
\begin{equation}
\Gamma_{j\bar{\jmath}}\rightarrow\Gamma^\prime_{j\bar{\jmath}}=
(M\Gamma M^T)_{j\bar{\jmath}}. \label{eq8.4}
\end{equation}
The matrix $M$ is given by (\ref{15}), and belongs to the set of
matrices that constitute the five-dimensional representation of the
permutation group of three elements. This group is generated by the
matrices
\begin{equation}
\begin{array}{cc} \left( \begin{array}{ccccc} \frac{1}{5} &
\frac{\sqrt{3}}{5} & \frac{1}{\sqrt{5}} & \frac{\sqrt{7}}{5} & \frac{3}{5} \\
\frac{\sqrt{3}}{5} & \frac{1}{2} & \frac{\sqrt{3}}{2\sqrt{5}} & 0 & -
\frac{2\sqrt{3}}{5} \\ \frac{1}{\sqrt{5}} & \frac{\sqrt{3}}{2\sqrt{5}} & -
\frac{3}{14} & -\frac{4}{\sqrt{35}} & \frac{6}{7\sqrt{5}} \\\frac{\sqrt{7}}{5}
& 0 &  - \frac{4}{\sqrt{35}} & \frac{1}{2} & -\frac{3}{10\sqrt{7}} \\
\frac{3}{5} & -\frac{2\sqrt{3}}{5} & \frac{6}{7\sqrt{5}} & -
\frac{3}{10\sqrt{7}} & \frac{1}{70} \end{array} \right), & \left(
\begin{array}{ccccc} \;1\;&&&& \\ &\;-1\;&&& \\ &&\;1\;&& \\ &&&\;-1\;& \\
&&&&\;1\; \end{array} \right) \end{array}.
\end{equation}
It is straightforward to verify, by changing the integration variable
as
\begin{equation}
\eta \rightarrow \eta^\prime = \eta,1-\eta,\frac{\eta-1}{\eta},\frac{\eta}
{\eta-1},\frac{1}{1-\eta},\frac{1}{\eta},\label{eq8.6}
\end{equation}
that $\Gamma_{j\bar{\jmath}}$ is invariant under (\ref{eq8.4}).
Furthermore, by replacing
\begin{equation}
\eta \rightarrow \bar{\eta}, \label{eq8.7}
\end{equation}
we see that
\begin{equation}
\Gamma_{j\bar{\jmath}} = \Gamma_{\bar{\jmath}j}. \label{eq8.8}
\end{equation}

The integrals (\ref{eq8.3}) are all divergent and need to be
regularized. In doing this we certainly wish to respect the symmetries
(\ref{eq8.4}) and (\ref{eq8.8}). If we introduce the (possibly
infinite) constants
\begin{eqnarray}
&& \alpha = \int \frac{d^2 \eta}{\pi} \nonumber\\ && \beta = \int
\frac{d^2 \eta}{\pi} \; \eta^{-1} \label{eq8.9}\\ && \gamma = \int
\frac{d^2 \eta}{\pi} \; \eta^{-2} \nonumber\\ && \delta = \frac{1}{2}
\int \frac{d^2 \eta}{\pi} \; \eta^{-1} \bar{\eta}^{-1},
\nonumber
\end{eqnarray}
we may calculate all other integrals we need by changing integration
variable as in (\ref{eq8.6}) and (\ref{eq8.7}). The result for the
amplitude $\Gamma_{j\bar{\jmath}}$ is
\begin{equation}
\begin{array}{rrrrr}
\Gamma_{00} =&27\alpha&-48\beta&+22\gamma&+18\delta\nonumber\\
\Gamma_{11} =&6\alpha&-90\beta&-6\gamma&+\frac{225}{2}\delta\nonumber\\
\Gamma_{22} =&10\alpha&+30\beta&+10\gamma&+\frac{45}{2}\delta\nonumber\\
\Gamma_{33} =&14\alpha&&-14\gamma&\nonumber\\
\Gamma_{44} =&18\alpha&-72\beta&+18\gamma&+72\delta\label{eq8.10}\\
\Gamma_{20} = \Gamma_{02} =&2\sqrt{5}\alpha&-6\sqrt{5}\beta&+12\sqrt{5}\gamma&
+9\sqrt{5}\delta\nonumber\\
\Gamma_{40} = \Gamma_{04}=&6\alpha&+66\beta&+36\gamma&
-36\delta\nonumber\\
\Gamma_{31} = \Gamma_{13}=&2\sqrt{21}\alpha&-15\sqrt{21}\beta&-2\sqrt{21}
\gamma&\nonumber\\
\Gamma_{42} = \Gamma_{24}=&6\sqrt{5}\alpha&-3\sqrt{5}\beta&+6\sqrt{5}\gamma&
-18\sqrt{5}\delta,\nonumber
\end{array}
\end{equation}
with all other components vanishing.

To regularize the integrals (\ref{eq8.9}) we modify the integration measure as
\begin{equation}
\int \frac{d^2\eta}{\pi} \rightarrow \int \frac{d^2 \eta}{\pi} \;
\lambda(\eta),
\end{equation}
where $\lambda(\eta)$ should be invariant under (\ref{eq8.6}) and
(\ref{eq8.7}), and go to $1$ as the regulator is turned off. We will
propose two different choices for $\lambda(\eta)$. The first is
\begin{eqnarray}
\lambda(\eta) &=& \frac{1}{6} ( |\eta|^{2 \epsilon_1} |1-\eta|^{2 \epsilon_2}
 \label{102}\\
&& + {\rm five\;\;terms\;\;with\;\;\eta\;\;replaced\;\;as\;\;in\;\;}
(\ref{eq8.6})) \nonumber.
\end{eqnarray}
The integrals could now be calculated using the formula
\cite{Gross-Harvey-Martinec-Rohm}
\begin{eqnarray}
&& \int \frac{d^2 \eta}{\pi} \; |\eta|^\alpha |1-\eta|^\beta \eta^n
(1-\eta)^m \\ && = (-1)^{n+m} \frac{
\Gamma(-1-\frac{\alpha}{2}-\frac{\beta}{2})
\Gamma(1+n+\frac{\alpha}{2}) \Gamma(1+m+\frac{\beta}{2})}
{\Gamma(-\frac{\alpha}{2}) \Gamma(-\frac{\beta}{2})
\Gamma(2+n+m+\frac{\alpha} {2}+\frac{\beta}{2})}, \nonumber
\end{eqnarray}
which converges for $Re(\alpha+\beta+n+m+2)<0$, $Re(\alpha+n)>-2$,
$Re(\beta+m)>-2$, and is to be understood in the sense of analytic
continuation elsewhere. We get
\begin{eqnarray}
&&\alpha = -\frac{1}{12} \frac{(\epsilon_1-\epsilon_2)^2}{\epsilon_1 +
\epsilon_2}+\frac{1}{3}\frac{\epsilon_2^2}{\epsilon_1} +
\frac{1}{3}\frac{\epsilon_1^2}{\epsilon_2}\nonumber\\ &&\beta =
\frac{1}{2} \\ &&\gamma = 0 \nonumber\\ &&\delta = \frac{1}{3}
\frac{1}{\epsilon_1}+\frac{1}{3} \frac{1}{\epsilon_2} -
\frac{1}{3}\frac{1}{\epsilon_1+\epsilon_2} \nonumber
\end{eqnarray}
modulo terms which vanish as $(\epsilon_1,\epsilon_2)\rightarrow
(0,0)$. If we furthermore put $\epsilon_1=\epsilon_2=\epsilon$ the
result is
\begin{equation}
\alpha = 0 \;\;\;,\;\;\; \beta = \frac{1}{2} \;\;\;,\;\;\; \gamma = 0
\;\;\;,\;\;\; \delta = \frac{1}{2} \frac{1}{\epsilon}. \label{eq8.15}
\end{equation}
Our second regulator is
\begin{eqnarray}
\lambda(\eta) & = & \theta(|\eta|-\epsilon) \\
&& \times ( {\rm five\;\;factors\;\;with\;\;\eta\;\;replaced\;\;as\;\;in\;\;}
(\ref{eq8.6})), \nonumber
\end{eqnarray}
where $\theta$ denotes the step function. Now we get
\begin{equation}
\alpha = \frac{1}{\epsilon^2}-\frac{2}{\pi} \frac{1}{\epsilon} \;\;\;,\;\;\;
\beta = \frac{1}{2} \;\;\;,\;\;\; \gamma = 0 \;\;\;,\;\;\; \delta = - 4 \ln
\epsilon
\end{equation}
modulo terms which vanish as $\epsilon \rightarrow 0$. This latter
regularization will prove less useful, though, since it fails to
eliminate the divergence of $\alpha$.

We should now compare the string amplitudes found in this section with
the field theory results of section~7. From our discussions in
section~7 we expect the $s$-channel contribution from exchange of the
scalar particle to be
\begin{equation}
(\Gamma^{(s)}_{\sigma})_{j\bar{\jmath}} \sim \delta_{j\bar{\jmath}}
(j(j+1)-6)^{-1},
\end{equation}
whereas the contribution $(\Gamma^{(s)}_{A})_{j\bar{\jmath}}$ from
tensor particle exchange should vanish for $|j-\bar{\jmath}|>2$ and
might diverge for $j=\bar{\jmath}=1$ since this representation is
``on-shell''. If we denote the total $s$-channel contribution as
$\Gamma^{(s)} = \Gamma^{(s)}_{\sigma} + \Gamma^{(s)}_{A}$ and
calculate the complete amplitude, including $t$- and $u$-channel
contributions, we get a result proportional to (\ref{eq8.10}) with the
constants $\alpha$, $\beta$, $\gamma$ and $\delta$ given by
\begin{equation}
\left( \begin{array}{c} \alpha \\ \beta \\ \gamma \\ \delta \end{array}
\right) = \left( \begin{array}{cccccccc} \!588\!&0&720&\!1050\!&\!162\!&
\!-336\sqrt{5}\!&0&192\sqrt{5}\\ 0&0&240&210&\!-30\!&-84\sqrt{5}&
\!-70\sqrt{5}\!&-62\sqrt{5}\\ 0&0&\!-240\!&\!-525\!&\!135\!&336\sqrt{5}&0&
48\sqrt{5}\\ 0&\!196\!&180&84&\!100\!&0&\!-56\sqrt{5}\!&\!-120\sqrt{5}\!
\end{array} \right) \left( \begin{array}{c} \Gamma^{(s)}_{00} \\
\Gamma^{(s)}_{11} \\ \Gamma^{(s)}_{22} \\ \Gamma^{(s)}_{33} \\
\Gamma^{(s)}_{44} \\ \Gamma^{(s)}_{02} \\ \Gamma^{(s)}_{13} \\
\Gamma^{(s)}_{24} \end{array} \right).
\end{equation}
Our result (\ref{eq8.15}) seems to indicate that all
$\Gamma^{(s)}_{j\bar{\jmath}}$ components are finite except
$\Gamma^{(s)}_{11}$. We therefore conclude that the effective field
theory corresponding to our string theory has no $\phi^3 (x)$
coupling, since otherwise $\Gamma^{(s)}_{22}$ would be divergent. The
divergence of $\Gamma^{(s)}_{11}$ is of course due to the
$A_{\mu\nu}(x)$ resonance. However, our example is to small to allow
for a determination of the coupling constants in the effective quantum
field theory.

Finally, let us consider the question whether the amplitude is dual in
the sense of section 4. More precisely, is it possible that the
$s$-channel contribution $\Gamma^{(s)}$ in itself is invariant under
(\ref{eq8.4}) and (\ref{eq8.8})? This means that $\Gamma^{(s)}$ should
have the form (\ref{eq8.10}) with the additional requirement that
$\Gamma^{(s)}_{04} = \Gamma^{(s)}_{40} =0$ since
$\Gamma^{(s)}_{j\bar{\jmath}}=0$ for $|j-\bar{\jmath}|>2$.  We see
that this is not possible if $\Gamma^{(s)}_{11}$ and/or
$\Gamma^{(s)}_{22}$ are infinite, since then other components of
$\Gamma^{(s)}$ would be infinite as well. We have thus found, that not
only must the $\phi^3(x)$ coupling be absent in a dual model, but the
$A_{\mu\nu}(x)$ field should couple in such a way to the $\phi(x)$
field that the $(j,\bar{\jmath})=(1,1)$ pole from the $A_{\mu \nu}$
propagator cancels.

{\vspace*{10mm} \large \bf \noindent 11. The continuum limit\\}
We have seen that the complexity of the problem appears to increase with
the spin $j_0$. There are reasons to expect simplifications, however,
as $j_0 \rightarrow \infty$. This limit means that the wavelength of
the incoming particles gets very small compared to the curvature
radius of the $SU(2)$ target space. It should therefore be a good
approximation to neglect the curvature altogether. We will not pursue
this approach in detail in this publication, but merely indicate what
techniques may be used.

Let us therefore introduce the variables $x=j/j_0$ and
$\bar{x}=\bar{\jmath}/j_0$, where $j$ and $\bar{\jmath}$ as before are
the spins of the left and right representations exchanged in a
four-point scattering respectively. We see that $0 \leq x,\bar{x} \leq
2$. As $j_0 \rightarrow \infty$ the spacing between the allowed values
of $x$ and $\bar{x}$ goes to zero. It is therefore natural to take a
continuum limit and replace the functions
$f_{j\bar{\jmath}}(\eta,\bar{\eta})$ by
$f(x,\bar{x};\eta,\bar{\eta})$. The matrices $A$ and $B$ of (\ref{7})
then turn into differential operators. We see that
\begin{equation}
(Af)(x) = A(x) f(x),
\end{equation}
and that
\begin{equation}
(Bf)(x) = \left( B_0(x) + B_1(x) \frac{\partial}{\partial x} + B_2(x)
\frac{\partial^2}{\partial x^2} \right) f(x) + {\cal
O}(\frac{1}{j_0^3}),
\end{equation}
where
\begin{eqnarray}
A(x) & = & x(x+\frac{1}{j_0})(1+\frac{1}{j_0})^{-1}-2 \nonumber\\
B_0(x) & = & (x^2-2) + \frac{1}{j_0} x(1-x) +
\frac{1}{j_0^2}(x^2-x+\frac{3}{16}-\frac{1}{4x^2}) \nonumber\\ B_1(x)
& = & \frac{1}{j_0^2} \frac{x}{2} \\ B_2(x) & = & \frac{1}{j_0^2}
(\frac{x^2}{4}-1). \nonumber
\end{eqnarray}
To lowest order in $1/j_0$, $A$ and $B$ commute, which means that the
continuum counterparts of the $\alpha$-matrices are diagonal, and we
may thus write the correlation function as a global product of a
holomorphic and an anti-holomorphic function, just as for a string in
flat space:
\begin{equation}
f(x,\bar{x};\eta,\bar{\eta}) = c(x) |\eta|^{2 A(x)} |1-\eta|^{- 2
B_0(x)}.
\end{equation}

To determine the function $c(x)$ we must go to the second order in
$1/j_0$ so that the $A$ and $B$ operators no longer commute. Our
differential equation is then
\begin{equation}
\frac{\partial f(x;\eta)}{\partial \eta} = \left( \eta^{-1} A(x) +
(1-\eta)^{-1}(B_0(x)+B_1(x)\frac{\partial}{\partial x}+B_2(x)
\frac{\partial^2}{\partial x^2}) \right) f(x;\eta) + {\cal O}(\frac{1}{j_0^3}).
\end{equation}
We see that as $\eta \rightarrow 0$, the general solution behaves as
\begin{equation}
f(x;\eta) = f_0(x) \eta^{A(x)} (1-\eta)^{-B_0(x)} (1+{\cal O}(\eta))
\end{equation}
for some function $f_0(x)$. As the solution is continued analytically
around a singular point, the function $f_0(x)$ undergoes a linear
transformation. We see that $f_0(x) \mapsto \exp(2\pi i A(x)) f_0(x)$
under analytic continuation around $\eta=0$. To determine the
behaviour under analytic continuation around $\eta=1$ it is convenient
to think of the $B$ operator as the sum of a ``free'' term $B_0(x)$
and an ``interaction term'' of order $1/j_0^2$. We then change
variables from $f(x;\eta)$ to $f_0(x;\eta)$ defined by
\begin{equation}
f(x;\eta) = f_0(x,\eta) \eta^{A(x)}(1-z)^{-B_0(x)}
\end{equation}
and treat the problem in the ``interaction picture''. The equation
(\ref{7}) now reads
\begin{eqnarray}
\frac{\partial f_0(x;\eta)}{\partial \eta} & = & (1-\eta)^{-1} \left( B_1(x)
\frac{\partial}{\partial x} + B_2(x) \frac{\partial^2}{\partial x^2} \right.
\nonumber\\
&& + \ln \eta \;
\left(A^\prime(x)B_1(x)+A^{\prime\prime}(x)B_2(x)+2A^\prime(x)B_2(x)
\frac{\partial}{\partial
x}\right) \nonumber\\ && - \ln (1-\eta) \;
\left(B_0^\prime(x)B_1(x)+B_0^{\prime\prime}(x)B_2(x)+2B_0^\prime(x)B_2(x)
\frac{\partial}{\partial
x}\right) \nonumber\\ && + \ln^2 \eta \; A^{\prime 2}(x)B_2(x) - 2 \ln
\eta \ln (1-\eta) \; A^\prime(x) B_0^\prime(x) B_2(x) \nonumber\\ && +
\left. \ln^2 (1-\eta) \; B_0^{\prime 2}(x)B_2(x) + {\cal
O}(\frac{1}{j_0^3}) \right) f_0(x;\eta).
\end{eqnarray}
The function $f_0(x;\eta)$ on the right hand side could be replaced by
$f_0(x)+{\cal O}(1/j_0^2)$. Analytic continuation along a contour $C$
thus transforms $f_0(x)$ as
\begin{equation}
f_0(x) \mapsto f_0(x)+\int_C d\eta \frac{\partial f_0(x;\eta)}{\partial \eta}.
\end{equation}
By evaluating these integrals, we may determine the counterpart of the
$\alpha$-matrices in (\ref{17a}). Finally, we may take the limit $j_0
\rightarrow \infty$, and solve condition (\ref{17}) for the function
$c(x)$.

{\vspace*{10mm} \large \bf \noindent 12.  Discussion\\}
The main result of this paper is the discussion of how quantum field theory
calculations on a group manifold could be performed. The possible
amplitudes depend on only a few arbitrary coupling constants. We have
also verified that the results from string theory and quantum field
theory agree for some simple examples. In the $k=4$ case, we were able
to extract some non-trivial information concerning the possibility of
constructing dual string amplitudes. A major obstacle for the
interpretation of the results has been that all information in the
amplitude is contained in only four divergent constants, the
calculation of which requires a somewhat arbitrary regularization
procedure. The two limits $k=4$ and $k \rightarrow \infty$, which we
have considered in this paper, have the advantage of being exactly
solvable, but for our purposes they are not really sufficient. We need
results from an intermediate regime, where the curvature of the target
space could be expected to play an important role, and the $SU(2)$
representations are big enough so that the result contains more
non-trivial information.

As we have already mentioned, the obvious way to continue the program
is to evaluate some string scattering amplitudes numerically, and thus
determine the coupling constants in the corresponding effective field
theory. A problem is that we may expect such amplitudes to be
divergent, when interpreted literaly, just as in the $k=4$ case. In
flat space string theory, we are used to continue analytically in the
external momenta to get a sensible answer, which is equivalent to what
we did for $k=4$ although our momenta are really discrete. Apart from
the practical problem of implementing analytical continuation
numerically, we may expect this procedure to be insufficient for
levels exceeding four, however. The reason is that, as we have already
noticed, the conformal field theory correlation functions are not
globaly a product of a holomorphic and an anti-holomorphic function
any longer. Consequently they have singularities with several
different exponents simultaneously as we let two insertion points
coalesce. The regulator (\ref{102}) will therefore in general fail to
produce a convergent integral. We expect these difficulties to be
tractable, though, and hope to come back to this issue shortly.

We have already mentioned the possibility of replacing the group
$SU(2)$ with its non-compact relative $SU(1,1)$. Not only does this
group provide us with a time direction, which raises interesting
questions concerning the unitarity of the theory, but more important
for our purposes is that it has a set of representations labeled by a
continuous variable $j$. This would allow us to impose the constraint
that scattering amplitudes should be analytic in $j$. Analyticity of
the $S$-matrix plays an important role in flat space. It might also
prove to be a convenient way of regularizing divergent amplitudes.
More general manifolds, with a non-vanishing dilation expectation
value, certainly also merit study. A suitable first step should be the
coset manifolds, among which is the two-dimensional black hole
solution to string theory \cite{Witten2}.

Our model may only be studied at tree level, since the total conformal
anomaly is different from zero. At the present stage, this is not a
serious problem, but eventually it would be interesting to consider
higher loop contributions in a fully consistent theory.

I would like to thank Lars Brink for numerous discussions and much
encouragement during the progress of the work. I have also benefited
from discussions with Stephen Hwang, Christian Preitschopf and Bo
Sundborg.

\end{document}